\documentclass{article}

\usepackage{arxiv}

\usepackage{graphicx}
\usepackage{amsmath, amssymb, bm}
\usepackage[FIGBOTCAP]{subfigure}
\usepackage{multirow}
\usepackage{mathtools}
\usepackage{enumitem}
\usepackage{bbold}
\usepackage{placeins}
\usepackage{ragged2e}
\usepackage{float}
\usepackage{lipsum}
\usepackage{cuted}

\title{Effect of Volume Fraction and Fiber Distribution on Stress Transfer in a Stochastic Framework of Continuous Fiber Composite: A Micromechanical Study}

\author{
 Ajinkya V. Sirsat \\
  MUSE Lab, Room no.206, Satish Dhawan Block\\
  Indian Institute of Technology Ropar\\
  Rupnagar-140001, Punjab,India \\
  \texttt{ajinkyavsirsat@gmail.com} \\
   \And
 Srikant S. Padhee\\
  MUSE Lab, Room no.206, Satish Dhawan Block\\
  Indian Institute of Technology Ropar\\
  Rupnagar-140001, Punjab,India \\
  \texttt{sspadhee@iitrpr.ac.in} \\
}

\begin{document}
\maketitle
\begin{abstract}
In fiber Reinforced Composites (FRC) fiber breakage is a common phenomenon resulting in stress concentration. This high stress gets transfer in the vicinity of the breakage which is quantified by Stress Transfer Coefficient (STC). In this paper, an attempt is made to check the effect of fiber volume fraction and the distribution of the fibers on STC and ineffective length. The fiber volume fraction is changed considering three cases: 1) by changing the number of fibers, 2) by changing the dimension of the Represntative Volume Element (RVE) and 3) by changing the fiber radius. Cases with change in dimension of RVE and change in fiber radius, periodic and semi-random arrangents of fibers are considered. From the analysis of 200 RVE's for each volume fraction in random and semi- random arrangements, it is observed that the distribution of STC does not follow any standard distribution, even if the fiber arrangement follows the normal distribution. The fiber cross-sectional dimension plays a critical role in regaining the broken fiber strength. The periodic arrangement of fibers can be said to be beneficial over the random arrangement considering the stress transfer from the broken fiber.
\end{abstract}


\section{Introduction}
Material selection is a very important and complex task for a design engineer. With the advancement in the technology, it is now becoming easy to analyse these material for a structure under different loading conditions. Nonetheless, the proper characterization of the material is a cumbersome task. The conventional materials are well characterized but the non-conventional or engineered materials like composite materials are still needed to be properly characterized. The tailoring of properties makes the composite find its application in many fields like automobile, aerospace, biomedical, etc.  Among all the composites the Fiber Reinforced Composite (FRC) is widely used. Hence the characterization of the FRC is important. Many researchers have tried different characterization methods to get different properties of FRC’s. Stress concentration in FRC is one of the parameters that need to be studied. Since stress concentration can cause very adverse effects in the structure. There are numerous reasons for the stress concentration in FRC, one of the reasons is the breakage of fiber. When a fiber break in the FRC structure, stresses in its vicinity increases. To maintain the equilibrium, this high stress gets transferred to other neighbouring fibers through the matrix~\cite{Hedgepeth1967}. The neighbouring fibers thus experience high stress which may lead to their breakage. This cumulative process of stress concentration and stress transfer continues which may lead to complete failure of the structure~\cite{yu_stress_2015}. This Phenomenon is quantified using the Stress Transfer Coefficient (STC).

The current study deals with the characterisation of FRC under fiber breakage condition, which is mostly dictated by the "micromechanics". While doing the micromechanical analysis it is not necessary to analyse the complete microstructure, but a part of it is considered which is known as Representative Volume Element (RVE). Being the representation of the composite, the RVE has the same effective mechanical properties as that of the complete macrostructure. Ideally, the fibers in the microstructure should be arranged periodically, but due to various errors during manufacturing, the fibers remain no longer periodic and are distributed randomly. In the literature, researchers have considered both types of fiber distribution i.e. periodic and random in the RVE to conduct the micromechanical analysis.

The analysis of the micromechanics after the fiber break in the FRC can be broadly classified as analytical and numerical. Both the analysis have vast literature showing that many researchers have contributed towards the characterization of the FRC. Cox~\cite{Cox1952} was the first to develop the shear lag model. This model is based on the basic assumption that the tensile load is carried only by the fibers while the matrix carries only shear stress. Using this model, the stress transfer from matrix to fiber (broken fiber regaining its strength) is predicted analytically. Hedgepeth~\cite{Hedgepeth1961} used the shear lag analysis to calculate the Stress Concentration Factor (SCF) which by definition is the same as the STC. The dynamic response factor (dynamic STC) was also calculated to get the insight of stress transfer during the transient phase. Hedgepeth and Dyke~\cite{Hedgepeth1967} further improved the model to remove the assumptions, elastic behaviour of the matrix, and 2D consideration of the model. SCF was calculated considering the behaviour of the matrix to be ideally plastic. Piggott~\cite{Piggott1966} also tried to get the analytical solution considering the plastic behaviour of the matrix.

Rosen~\cite{Rosen1964} presented an experimentally validated analytical solution to demonstrate the statistical distribution of the flaws or imperfections to be the cause of the fiber failure under the applied load. Smith~\cite{Smith1983} attempted using the shear lag model, to get the relation between the random SFC and the random spacing of the fibers. A correlation was also predicted between two consecutive fibers. Fukuda~\cite{Fukuda1985} extended Smith’s work, eliminating the assumption of rigid fiber, and presented a method for the derivation of statistical SCF using shear lag analysis. It was observed that, even if the fibers are randomly spaced based on Weibull distribution, the SCF distribution does not fit Weibull distribution or normal distribution. McCartney~\cite{McCartney1989} had done a thorough study to develop an analytical model for the micromechanical analysis of the stress transfer around fiber break or matrix crack, considering aspects like the residual thermal stresses, mismatch of Poisson ratio, bonding of the interface (perfect bonding and slip at the interface).

Handling the approximations of the shear lag model Zhao and Ji~\cite{Zhao1997} presented an improved shear lag model to calculate the tensile stresses in the fibers, shear stresses in the matrix, and tensile and shear stresses at the interface. Nairn~\cite{Nairn1997} revisited the shear lag model to check its accuracy. Four assumptions were made to re-derive the shear lag model with a modification in the shear lag parameter ($\beta$) compared to the one derived by Cox~\cite{Cox1952}. The developed shear lag model proved to predict the average axial stresses in the fibers and the strain energy within the $20\%$ error. However, the model failed to predict the shear stress and the energy release rate. A shear lag model was developed by considering the elastic/plastic behaviour of the matrix by Landis and McMeeking~\cite{Landis1999}. J2 flow rule governing the matrix deformation of the matrix was used to derive the solution. Afonso and Ranalli~\cite{CarlosAfonso2005} combined the modified shear lag model and the methods of cell to develop a new model. Using this model effective properties of the composite were predicted. The key feature of the developed model was that it does not depend on the microstructure i.e. the model works effectively for any kind of fiber configurations.

Various theoretical models discussed above are the attempt, to study the physics governing the composite behaviour having broken fibers under applied axial load. These models have however limitations. All these models use assumptions to prove their validity. The analytical models generally are not useful in predicting the stress distribution for low fiber volume fractions and low fiber/matrix stiffness ratios~\cite{Nairn1997,Xia2002}. The models are generally developed for a particular case so their feasibility for other cases is limited.

Another approach used for the micromechanical analysis is by using the numerical methods (most common is the Finite Element Method (FEM)). The researchers have considered different parameters to study their effects on stress transfer. Chen~\cite{Chen1971} considered the perturbation effect and distortional energy criteria and with the use of FEM predicted that the composite with discontinuous fiber, regains up to $50$ to more than $80\%$ of the strength if the aspect ratio of the fiber exceeds the critical. Goree and Gross~\cite{Goree1980} using the numerical method, showed that in contrast to the shear lag solution matrix do carry tensile stresses of a significant amount while the shear stresses are $50\%$ less. The effect of fiber spacing on the stresses was also studied. It was found that the stresses in the fibers are the same as predicted by the shear lag model.

Nedele and Wisnom~\cite{Nedele1994} used the periodic hexagonal distribution of fibers with different volume fractions. It was inferred that the stress concentration at the fiber break has a very small influence on the other fibers to make them fail in tension. Heuvel et al.~\cite{VanDenHeuvel1998} studied RVE with the periodic square arrangement of fibers considering two cases (a) perfect fiber/matrix bond (b) poor fiber/matrix bond. It was validated through the experiment that the SCF increases with (i) increase in the fiber/matrix stiffness ratio, (ii) decrease in the interfiber spacing, and (iii) increase in the matrix yield stress. For transverse loading, Oh et al.~\cite{oh_interfacial_2006} studied the effect of volume fraction and the interfiber spacing for both periodic and random arrangements of fibers in the RVE’s. It was also observed that the random RVE must be used to predict the behaviour of the composite at the microscale. The same observation was made by Trias et al.~\cite{trias_random_2006} while comparing periodic and random models. The periodic model was suggested to calculate the effective properties of the composite.

The effect of randomness, volume fraction, and fiber-matrix stiffness ratio on the STC and the ineffective length was studied by Swolf et al.~\cite{Swolfs2013} and Yu et al.~\cite{yu_stress_2015}. Ganesh et al.~\cite{Ganesh2017} used a 2D planar array of fibers to study the dynamic effects after fiber breakage. The effect of number of broken fiber on stress transfer was studied by Piere et al.~\cite{St-Pierre2017}. The SFC and the ineffective length increases with the number of broken fibers. The critical number of broken fibers in the cluster is found to be equal to 16. Recently Barzegar et al.~\cite{Barzegar2020} have studied the progressive failure of fiber to predict the stress redistribution right from the first fiber break. Using various parameters like fiber volume fraction, random distribution of fiber strength, interface properties, and fiber/matrix stiffness ratio and by performing dynamic analysis, fiber/matrix debonding, matrix deformation, and curing residual stress were predicted. 

From the above literature, it is found that none of the researchers have tried to compare the different ways in which the volume fraction can be varied and its effect on STC and ineffective length. Also, a compromise between complete randomness and complete periodicity in fiber arrangement i.e. the semi-random arrangement of fibers in the RVE is not studied. The literature on the effect of randomness on the STC does not consider a large set of data to analyse completely the effect of fiber position on the stress transfer. In this paper, an attempt is made to study the effect of volume fraction and fiber distribution on the STC and the ineffective length using Finite Element Analysis (FEA). Three cases of variation of volume fraction are considered Case 1: by changing the number of fibers, Case 2: by changing the dimension of the RVE (Matrix volume fraction) and Case 3: by changing the fiber radius. A wide variation in the fiber volume fraction is applied $(10 ~\text{to}~70\%)$. In Case 1 the fiber arrangement is random. For Case 2 and Case 3 two types of fiber arrangements were considered, semi-random and periodic. 200 RVE’s for each fiber volume fraction, in the random and semi-random arrangements of fibers were analysed to get a large set of data. Based on the data an attempt is made to do a statistical study on the distribution of the STC in the random and semi-random fiber arrangement. A comparison study is performed among the three cases of the volume fraction variation.

This paper is further divided into three sections. In Section~\ref{sec:FEM} the finite element model is discussed in brief for the three cases, to carry out the desired study. Section~\ref{sec:R&D} discuss the results obtained by solving the FEM using the commercial software ABAQUS. Lastly, section~\ref{sec:conclusion} concludes the paper by summing up the findings of the results.

\section{Model Setup}\label{sec:FEM}
The volume fraction of fiber is defined as the ratio of the volume of total fibers to the total volume of the RVE. Mathematically it can be written as
\begin{equation}
	V_f = \frac{n\pi R_f^2 l}{D^2L};
\end{equation}
Where, $n$ are the number of fibers, $R_f$ is the radius of the fiber, $l$ is the length of fibers, $L$ is the length of RVE, and $D$ is crossectional size of the RVE. For present study, which deals with continuous fibers, as the length of fibers ($l$) and length of the RVE ($L$) are same, the above expression of volume fraction simplifies to:
\begin{equation}
	V_f = \frac{n\pi R_f^2 }{D^2} 
\end{equation}
From the above expression it can be observed that $n$, $R_f$, and $D$  influence the volume fraction and in turn influence the associated properties. The randomeness in these parameter influences  the overall response of composite, which in turn reflects the stochastic nature of the FRC. A deeper study is essential to quantify the influence of these  parameters. The present work, following three cases are considered:
\begin{itemize}
	\item [\textbf{Case 1:}] The dimension of the RVE ($D$) and the radius of the fiber ($R_f$) are kept constant while the number of fibers ($n$) is altered to change the volume fraction.
	\item [\textbf{Case 2:}] In this case, the volume fraction($V_f$) is varied by changing the dimension of the RVE ($D$) while keeping the number of fibers ($n$) and the radius of the fiber ($R_f$) constant.
	\item[\textbf{Case 3:}] In this case the dimension of the RVE ($D$) and  the number of fibers ($n$) are kept constant while the radius of the fiber ($R_f$) is varied to change the volume fraction.
\end{itemize}

Table~\ref{tab:tab1} shows the variation of the parameters in the respective cases.

\begin{table}[t]
	\caption{Variation of number of fibers, Dimension of the RVE ($D$) and fiber radius ($R_f$) with the Volume fraction ($V_f$) for the three cases respectively. For Case 2 and Case 3 the variation of the standard deviation of the fiber arrangement ($\sigma$) along with the Volume fraction ($V_f$) }
	\begin{center}
		\label{tab:tab1}
		\begin{tabular}{c c c c c c}
			& & & & & \\
			
			\hline
			\noalign{\vspace {.2cm}} 
			$V_f$ ($\%$) & Case 1 & \multicolumn{2}{c}{Case 2} & \multicolumn{2}{c}{Case 3} \\
			\noalign{\vspace {.2cm}}
			&Number of fibers& $D~(\mu m)$& $\sigma$ ($\mu m$)& $R_f~(\mu m)$& $\sigma$ ($\mu m$)\\
			\hline
			\noalign{\vspace {.2cm}} 
			10 & 4 & 70 & 1.14166 & $-$ & $-$ \\
			20 & 7  & 50 & 0.75 & 1.26 & 0.32\\
			30 & 11 & 40.45 & 0.43167 & 1.5 & 0.24\\
			40 & 14  & 35  & 0.25 & 1.8 & 0.014\\
			50 & 18 & 31.2 & 0.12667 & 2.0 & 0.0733\\
			60 & 21 & 28.6 & 0.03667 & 2.2 & 0.0066\\
			70 & 25 & 26.5 & 0.033 & $-$ & $-$\\
			\noalign{\vspace {.2cm}} 
			\hline
			
		\end{tabular}
	\end{center}
\end{table}

\subsection{Generation of RVE with random microstructure}\label{sec:RVE}
In literature random microstuctue is generated either through (a) using image processing based techniques,  or (b) using an algorithm. 
In the image processing technique  first a realistic microstructure image is obtained of an FRC using a characterization technique e.g. Scanning Electron Microscope (SEM). Then this image is processed to get the location of fibers. The data of fiber location is then used to generate Statistically Equivalent Representative Volume Element (SERVE) using any of the predefined algorithms~\cite{Wang2016,Parambil2016}.

The algorithm base RVE generation technique is more preferred as it enables generation of large number of representative sets.  Feder~\cite{Feder1980} developed the Random Sequential Algorithm (RSA) in which a jamming limit of $54.7\%$ was observed. Wongsto et al.~\cite{Wongsto2005} have achieved higher volume fraction RVE by disturbing the hexagonally arranged fibers. Around 500 non-overlapping fibers with $65\%$ of volume fraction were successfully arranged randomly by Melro et al.~\cite{melro_generation_2008}, based on a self-developed 3 step algorithm. Yang et al.~\cite{Yang2013} developed a Random Sequential Expansion (RSE) algorithm with which high volume fraction RVE is possible. Zhang and Yan~\cite{zhang_comparison_2017} developed another method in which two algorithms (i) random disturbance and (ii) perfect elastic collision were used. With this method, an RVE with periodically arranged fibers was converted to RVE with randomly distributed fibers by giving each fiber initial unit velocity in a random direction. Bheemreddy et al.~\cite{Bheemreddy2016} developed a three-step procedure to generate RVE: (a) Generation of RVE with fibers arranged in a square array, (b) global crisscrossing and (c) subframe selection. Wang et. al.~\cite{Wang2011} developed the shaking algorithm to generate the 2D RVE. The algorithm consists of a two-step procedure, global criss-crossing, and local disturbing of fibers. High fiber volume fraction can be achieved using this algorithm. This algorithm is efficient in generating random microstructure with relative ease in implementation, and hence it is the inspiration behind the process of generating RVE's in Case 2 and 3.The following subsections illustrates the method of generating RVE's in the three cases.
\begin{figure} 
	\centering
	\subfigure[]{\includegraphics[width=1.45in]{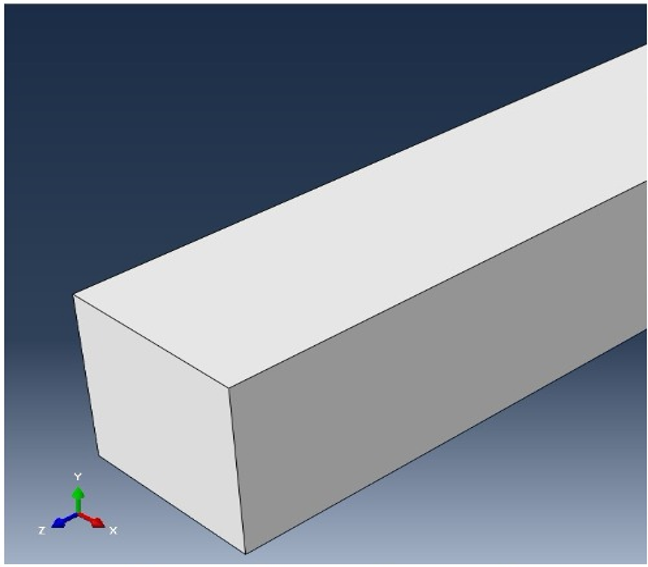}
	}
	~
	\subfigure[]{\includegraphics[width=1.45in]{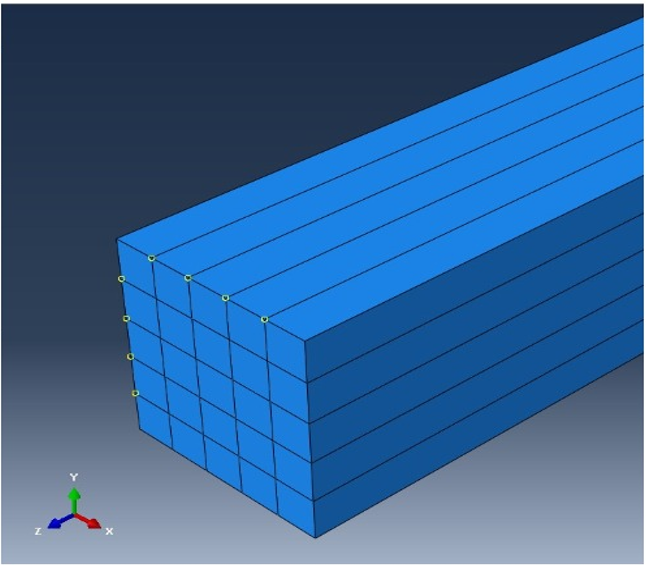}
	}
	~
	\subfigure[]{\includegraphics[width=1.45in]{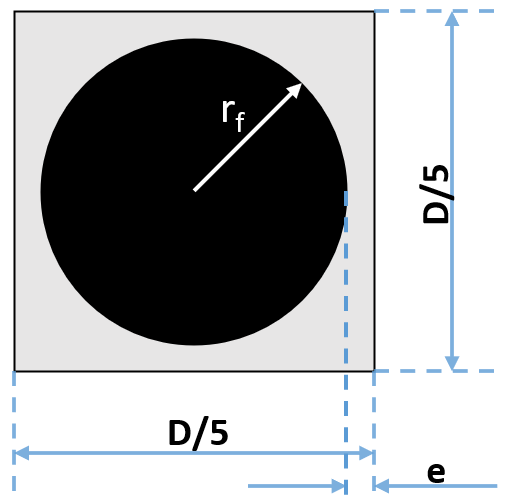}
	}
	\caption{Generation of RVE (a) Solid model of a block of size $D~\times~D~\times~100~\mu m$, (b)  Descritized block in 25 unit cells, (c)Typical Unit cell.}
	\label{fig:fig1}
\end{figure} 

\begin{figure*} 
	\centering
	\subfigure[]{\includegraphics[width=0.9in]{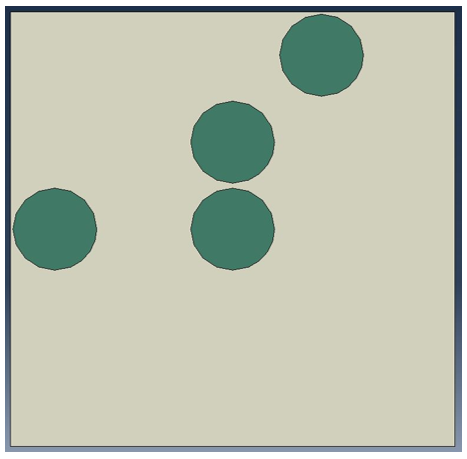}
	}
	~
	\subfigure[]{\includegraphics[width=0.9in]{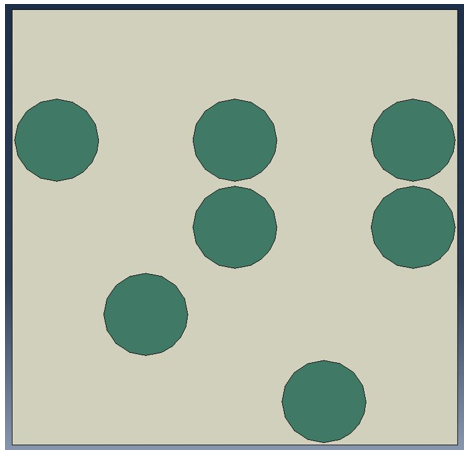}
	}
	~
	\subfigure[]{\includegraphics[width=0.9in]{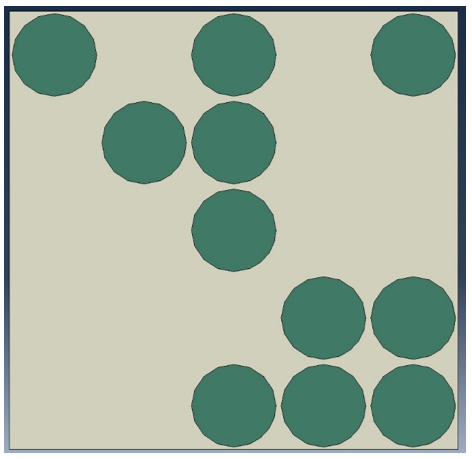}
	}
	~
	\subfigure[]{\includegraphics[width=0.9in]{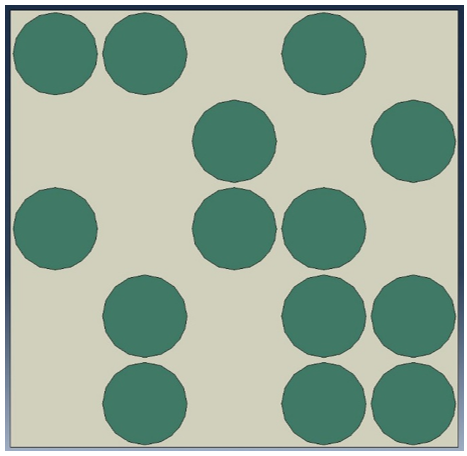}
	}
	~
	\subfigure[]{\includegraphics[width=0.9in]{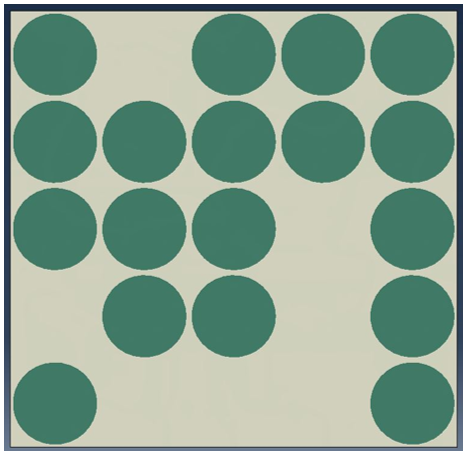}
	}
	~
	\subfigure[]{\includegraphics[width=0.9in]{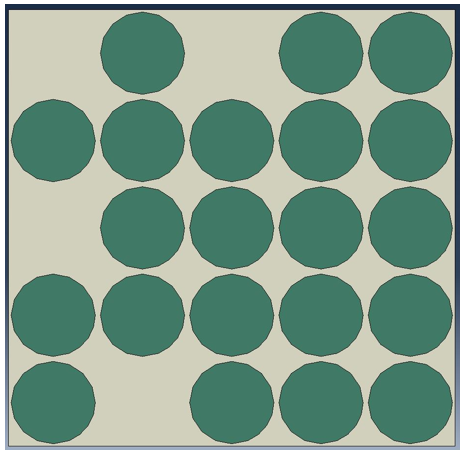}
	}
	\caption{One out of 200 RVE's for volume fractions: (a) $V_f$ $10\%$, (b)  $V_f$ $20\%$, (c)  $V_f$ $30\%$, (d)  $V_f$ $40\%$, (e)  $V_f$ $50\%$, (f)  $V_f$ $60\%$}
	\label{fig:fig2}
\end{figure*}
200 RVE's for each volume fraction in each case are analyzed. The reason for choosing the number of RVE's equal to 200 is discussed later in this section.
\subsubsection{Case 1:}~\label{sec:case1}
In this case the size of the RVE is $26.5 ~\times ~26.5 ~\times ~ 100~\mu m$. The radius of the fiber is $2.5\mu m$. To generate the complete RVE with random fiber arrangement following steps are performed:\noindent
\begin{itemize}
	\item [\textbf{Step 1}:] A block of the size corresponding to size of the RVE ($D$) and of length $100~\mu m$ is generated.Fig~\ref{fig:fig1}($a$)
	\item [\textbf{Step 2}:] The block is further divided into an array of 25 unit cells each of size $\frac{D}{5}~ \times  ~\frac{D}{5} ~\times  ~100$. Fig~\ref{fig:fig1}($b$) The centre of each of these unit cells makes the centre of the fibers. All RVE’s have the central fiber located at the geometrical centre of the RVE.This fiber is considered to be broken by applying the appropriate boundary condion.
	\item [\textbf{Step 3}:] Corresponding to the volume fraction, number of the centres excluding the central fiber (Table~\ref{tab:tab1}) are chosen randomly and cylinders resembling the fibers are generated with the radius $R_f$ and length $L$. Hence this case corresponds to complete randomness.
\end{itemize}

\noindent Fig.~\ref{fig:fig2} shows one of the RVE’s of each volume fraction. From the figure, it is clear that as the volume fraction increases the RVE becomes more periodic. 

\subsubsection{Case 2:}\label{sec:case2}
\begin{figure} 
	\centering
	\subfigure[]{\includegraphics[width=1.45in]{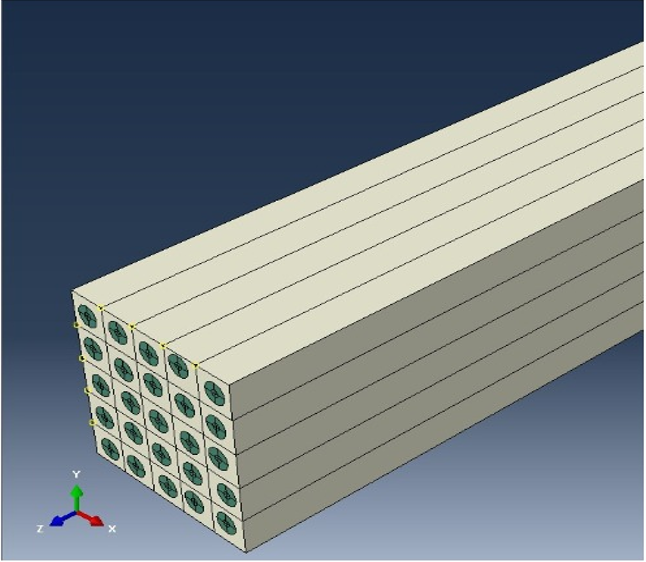}
	}
	~
	\subfigure[]{\includegraphics[width=1.45in]{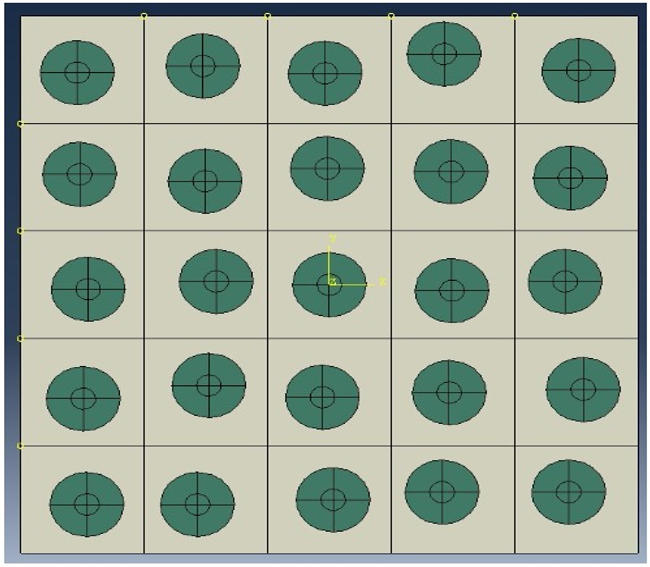}
	}
	\caption{A Typical RVE of Case 2 and 3 (a) Cross-sectional view of block with semi-random arrangement of fibers (complete RVE), (b) Isometric view of the generated RVE }
	\label{fig:fig3}
\end{figure}

Number of fibers, in this case, is $(n =)$ 25 and the radius is $2.5\mu m$. The variation of $D$ for volume fraction is shown in Table~\ref{tab:tab1}. The procedure for generation of RVE's is the same as that of Case 1 the difference being in the last step (Step 3). For this step two subcases are considered Semi-random and Complete Periodic. Fig~\ref{fig:fig3} shows Step 3 result. 
\paragraph{Semi-random:} 
In this case, the fibers are generated in each unit cell randomly following the normal distribution. The mean $(\mu)$ for the normal distribution is the centres of the unit cells. The standard deviation $(\sigma)$ for different volume fractions is shown in Table~\ref{tab:tab1}. In each RVE the positions of the fibers in the unit cells vary. Thus this case resembles the shaking model given by Wang et al. ~\cite{Wang2011}. From Table~\ref{tab:tab1} it is clear that with the increase in the volume fraction the standard deviation of fiber distribution decreases. This means as the volume fraction is increasing, $D$ is decreasing and hence the space for the fiber to move in a unit cell becomes less. Thus, for higher volume fractions the RVE tends to become periodic same as in Case 1. For volume fraction $70\%$ the RVE is pure periodic as there is no scope for fiber deviation from the mean position. The standard deviation is calculated as follows:
\begin{equation}
	\centering
	\sigma = \frac{\frac{D}{5}-2R_f-2e}{6}.
\end{equation}
Here, $2e$ is subtracted so that the fiber does not touch the boundaries of the unit cells, $e$ is the minimum distance from the boundary of the unit cell to the fiber periphery. The value of $e$ is taken to be equal to $0.25\mu m$. Fig~\ref{fig:fig1}c. Shows the cross-section of a typical unit cell.\footnote{The central broken fiber remains at its mean position in all the RVE’s.}

\paragraph{Complete Periodic:}
In this, the position of the fibers is exactly at the centres of the unit cell. The standard deviation ($\sigma$) is zero in this case.

\subsubsection{Case 3}\label{sec:case3}
$D$ for this case is taken to be $25\mu m$ and $n$ is kept 25. The values of $R_f$ for different volume fractions are shown in Table~\ref{tab:tab1}. Similar process as in Case 2, is used to generate RVE the only difference is instead of changing $D$, $R_f$ is varied. In this case also, two subcases similar to Case 2, were considered, Semi-random and Periodic. The std. deviation ($\sigma$) of fiber distribution in the semi-random part is as shown in Table~\ref{tab:tab1}. The value of $e$ in this case $0.28\mu m$.

\subsection{Material properties, BC and Meshing}~\label{sec:matprop}
The paper of Yu et al. ~\cite{yu_stress_2015} was referred for defining the material properties, boundary conditions and meshing of the fibers and matrix section in ABAQUS. Table~\ref{tab:tab4} shows the material properties of the fiber and matrix. Fibers are arranged in the '$z$' direction. To reproduce the microscopic state of the RVE in all three cases, a displacement of $0.1 \mu m$ is applied to RVE at $z = 0$ plane and $z-$symmetric boundary condition is applied to RVE at $z = ~100\mu m$ plane. The degree of freedom was released from the nodes of the central fiber to mimic the broken fiber in the real microstructure.

\begin{figure*} 
	\centering
	\subfigure[]{\includegraphics[width=2in]{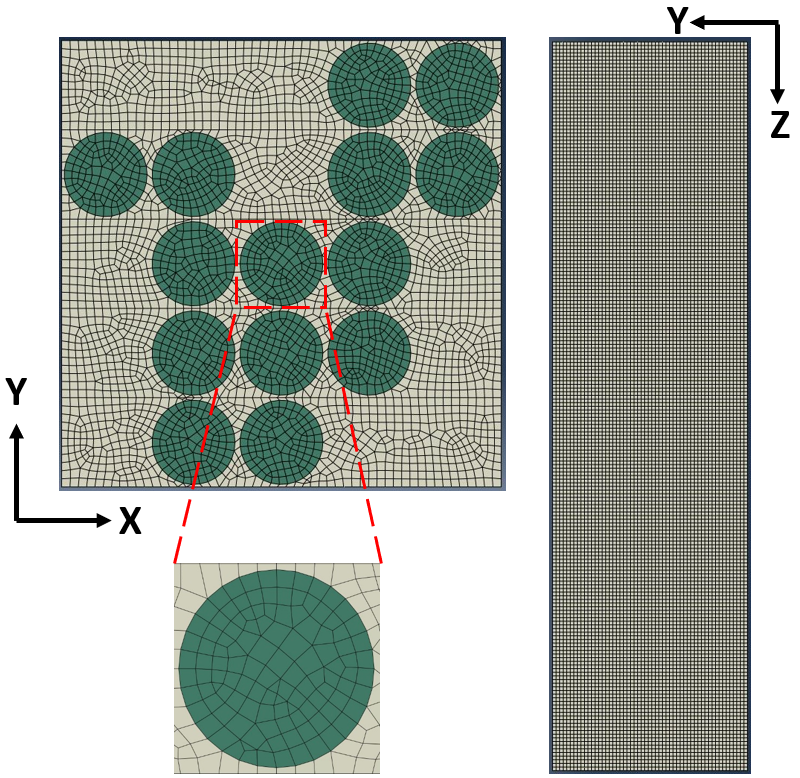}
	}
	~
	\subfigure[]{\includegraphics[width=1.9in]{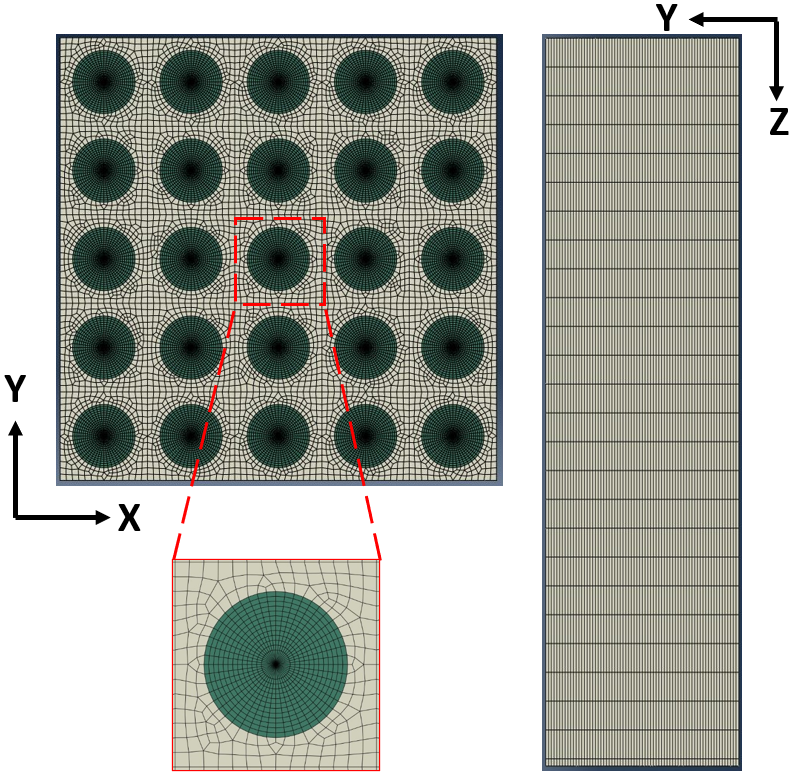}
	}
	~
	\subfigure[]{\includegraphics[width=1.9in]{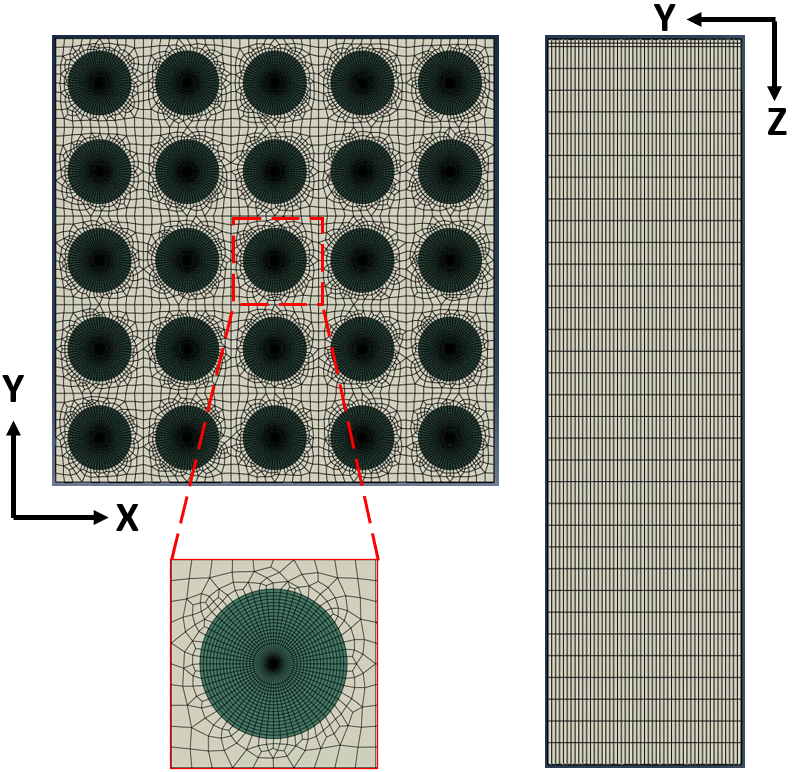}
	}
	\caption{Meshing in 1 of 200 RVE’s in all three cases (a) Case 1, (b) Case 2, (c) Case 3}
	\label{fig:fig4}
\end{figure*}

In Case 1, as the fibers were placed randomly, it was difficult to maintain a structured mesh in all the RVE’s. So, in Case 1, hex dominated meshing was used having linear brick element (C3D8R) and wedge element (C3D6) with an approximate global element size of $0.5 ~\mu m$ for meshing the RVE’s. In Case 2 and Case 3, linear brick type 8-node with hourglass control (C3D8R) element is used for meshing of the matrix, while fibers are meshed using C3D8R and C3D6 element type. For Case 2 the element size of a unit cell in $x$ and $y$ direction is taken to be equal to $0.5~\mu m$ and in the $z$ direction it is $4~\mu m$ up to breakage and in the breakage region, the element size is $1~\mu m$. To mesh the fiber it is partitioned to have an inner cell of radius $0.5~\mu m$. The inner cell only consists of wedge elements (C3D6) while the rest of the fiber is discretised using C3D8R elements. Along the circumference of the fiber including the inner cell, there are 48 nodes, while along the radius up to inner cells there are 13 nodes and in the inner cell, there are only two nodes along the radius.  In Case 3  the element size of the unit cell along the $x$ and $y$ direction is the same as Case 2 but the element size along the $z$ direction up to the fiber breakage region is $3~\mu m$ and in the fiber breakage region it is $0.5~\mu m$. For meshing the fiber similar technique was used as in Case 2 but having 68 number of nodes along the circumference and 17 number of nodes along the radius up to the inner cell. Fig~\ref{fig:fig4} shows the meshing in all three cases. 

\begin{table}
	\centering
	\caption{Fiber and matrix material properties}
	\label{tab:tab4}       
	\begin{tabular}{c c c}
		\hline\noalign{\smallskip}
		Material & Modulus (MPa)& Poisson's Ratio  \\
		\noalign{\smallskip}\hline\noalign{\smallskip}
		T800 Carbon Fibre & 394 & 0.3 \\
		Epoxy Resin & 3.4 & 0.4 \\
		\noalign{\smallskip}\hline
	\end{tabular}
\end{table}

\subsection{Convergence study}
To have reliable results it is essential to have convergence study both from the point of view of FEM and stochastic analysis. As the meshing and other FE based conditions are adapted from Yu et al.~\cite{yu_stress_2015} FEM based convergence study has not been carried out in this paper. However a convergence study based on number of iteration have been carried out. As the size of RVE is similar as that of the Yu et al.~\cite{yu_stress_2015}, the element size used in Case 1 assures the convergence of the results. Whereas, in the meshing of fibers in Case 2 and Case 3, as the element size used is less than the element size used by Yu et al.~\cite{yu_stress_2015} the convergence is hence guaranteed.  

\begin{figure}
	\centering
	\subfigure[]{\includegraphics[width=0.4\textwidth]{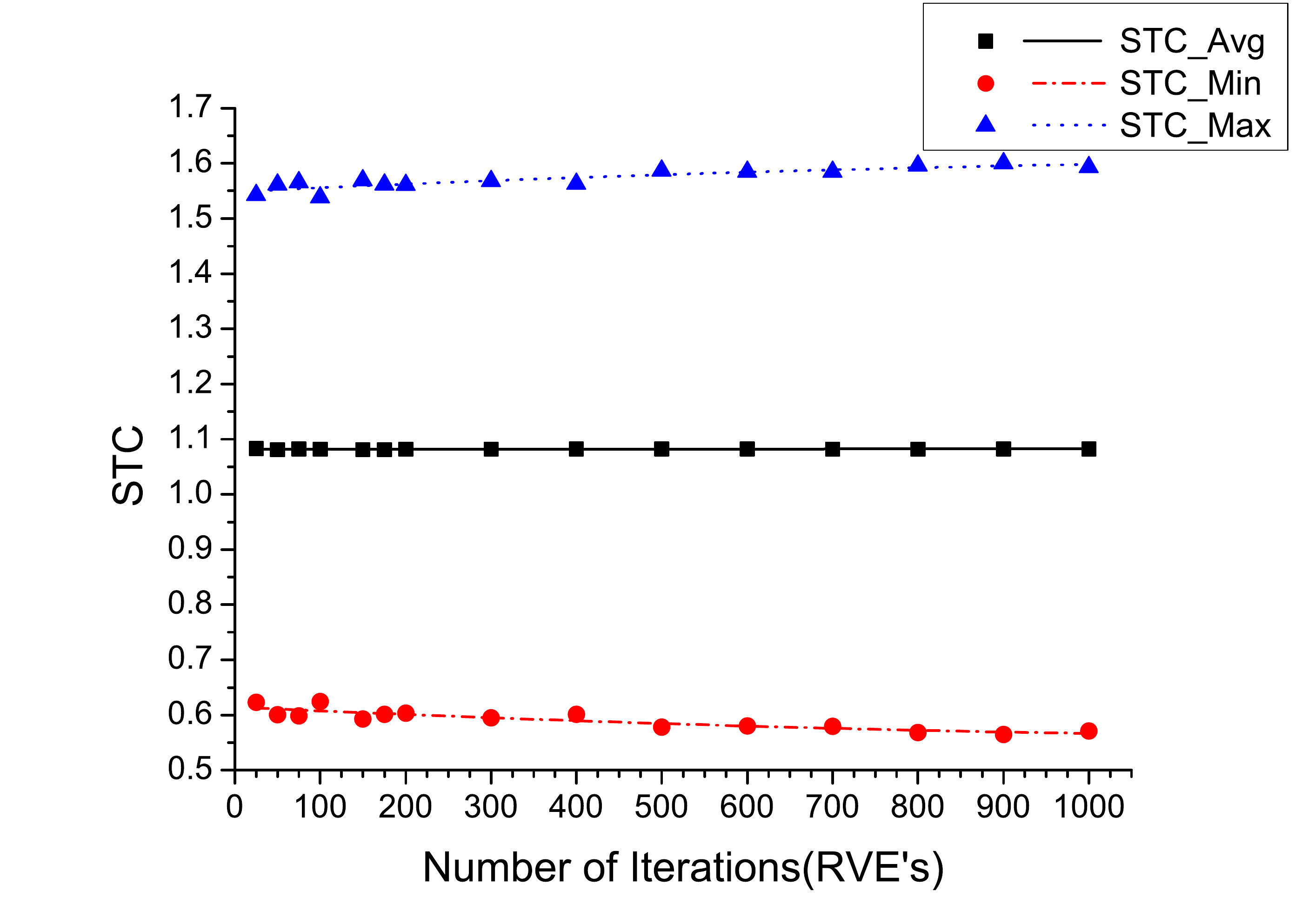}
		\label{fig:fig5a}
	}
	~
	\subfigure[]{\includegraphics[width=0.4\textwidth]{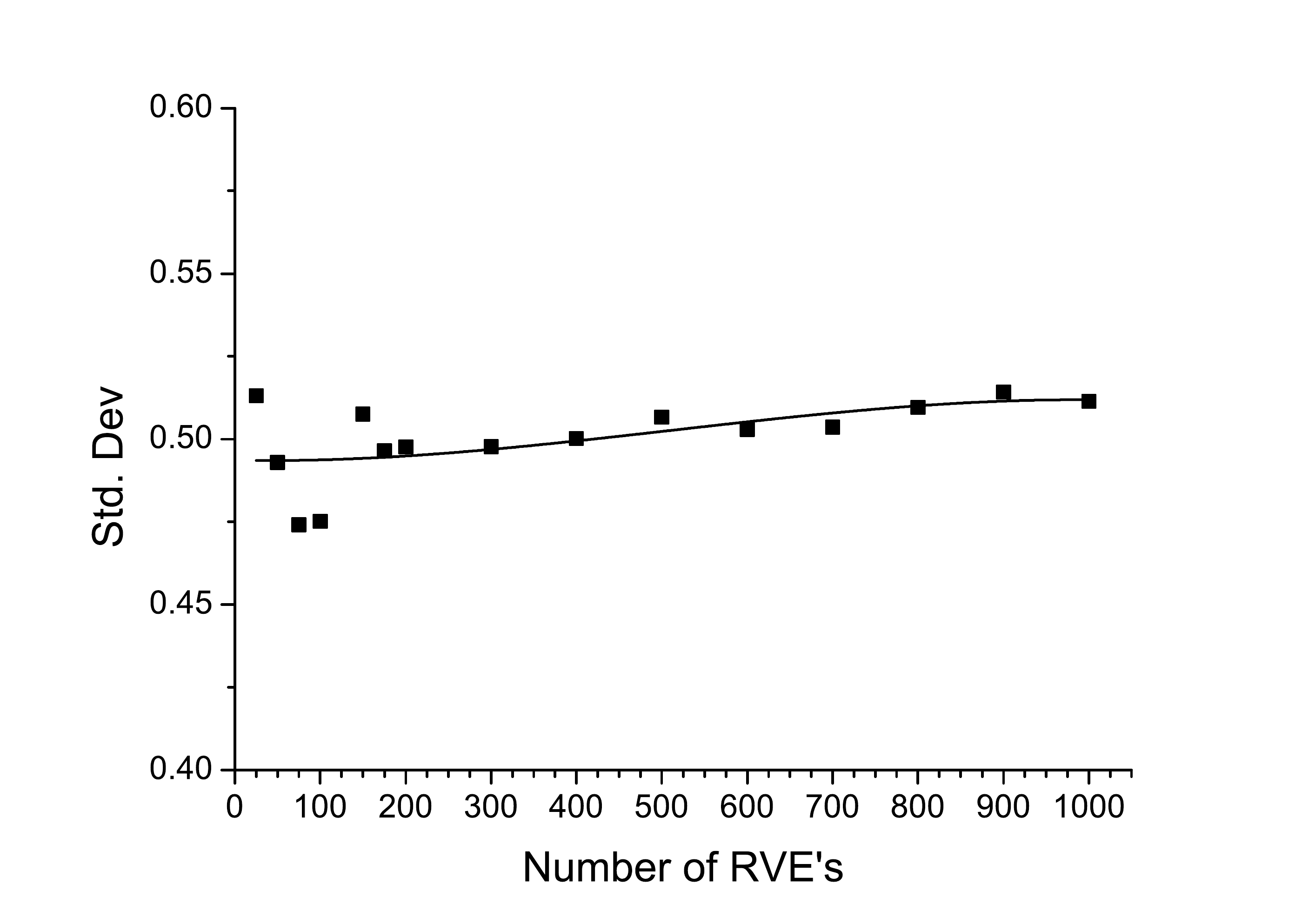}
		\label{fig:fig5b}
	}
	
	\caption{(a)Variation of average, maximum and minimum STC with the number of RVE's and, (b) Variation of the std. dev of the STC's for the respective number of RVE's}
	
\end{figure}

To predict the accurate behaviour of the effect of volume fraction on the STC in these three cases, a large set of data needs to be evaluated. For this large number of RVE's with different fiber arrangements for every volume fraction in the respective cases needs to be analysed. To find the optimal number of RVE's a convergence study for the stochastical analysis was carried out. The study includes analysis of upto 1000 number of RVE's of Case 2  having volume fraction 20 $\%$, whose average STC's are plotted against the corresponding number as shown in  Fig~\ref{fig:fig5a}. From the figure it can be observed that the average STC values for 200 RVE's and 1000 RVE's is almost the same. From  Fig~\ref{fig:fig5b} the variation of the standard deviation of the values of STC for various number of RVE's it can be seen that the standard deviation have a stable variation after 200 RVE's. From this result the optimum number of RVE's is chosen to be 200. The process of generating these 200 RVE’s for each volume fraction in each case was automated using Python script. The script includes all the information right from generating the model to exporting the desired result. 

\section{Results and Discussions}\label{sec:R&D}
The current study as mentioned earlier involves the study of the effect of volume fraction on STC and ineffective length and the study of effects of fiber distribution on STC and ineffective length. STC is defined as the ratio of the average stress near fiber breakage ($S^\star$) to average stress in the far field (S).
\begin{equation}
\centering
STC=  \frac{S^\star}{S}
\end{equation}
Ineffective length is the fiber length corresponding to 90 \% of the far field stress~\cite{Rosen1964}.

\subsection{Analysis of Case 1.}
\begin{figure} 
	\centering
	\subfigure[]{\includegraphics[width=0.4\textwidth]{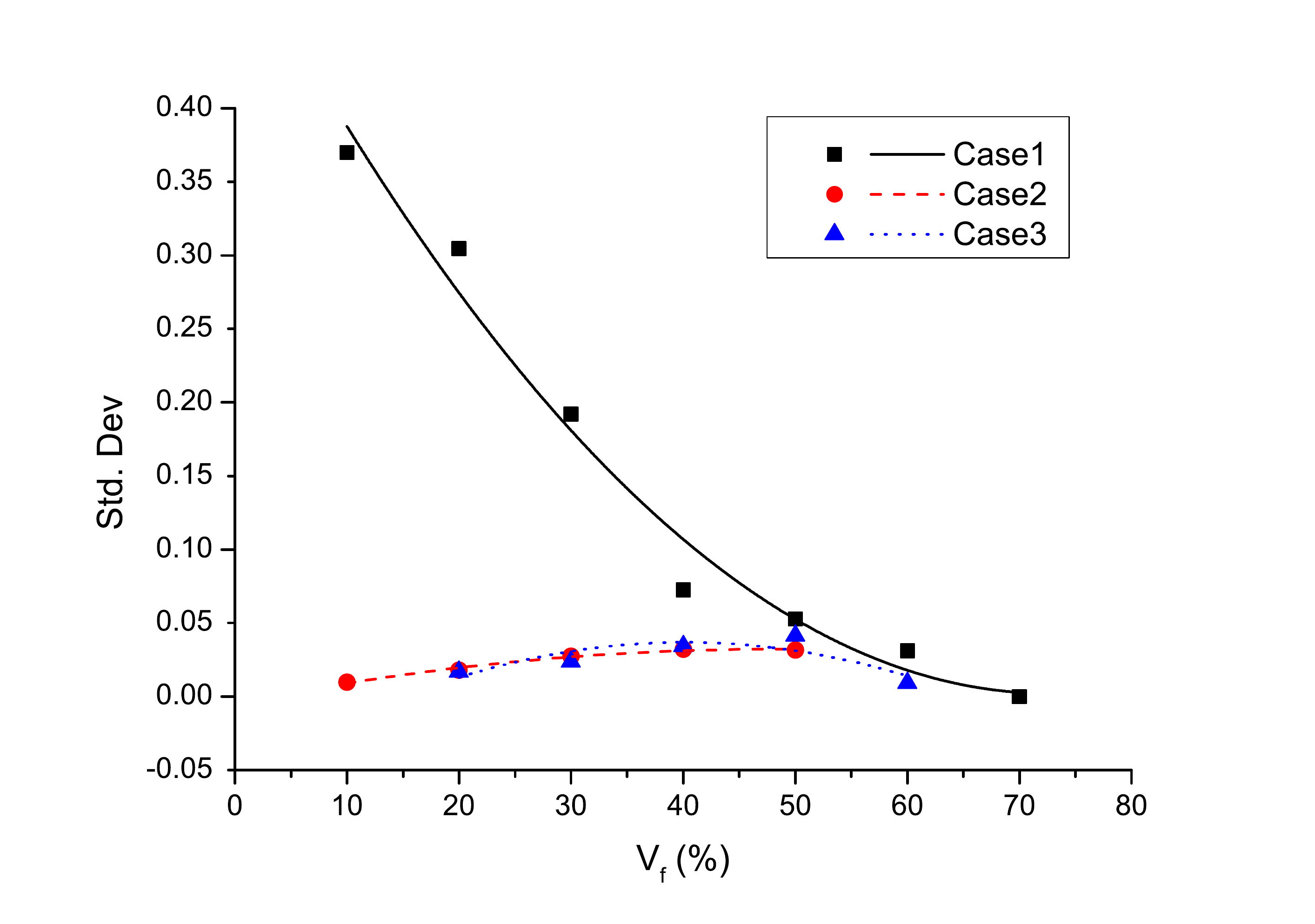}
		\label{fig:fig6a}
	}
	~
	\subfigure[]{\includegraphics[width=0.4\textwidth]{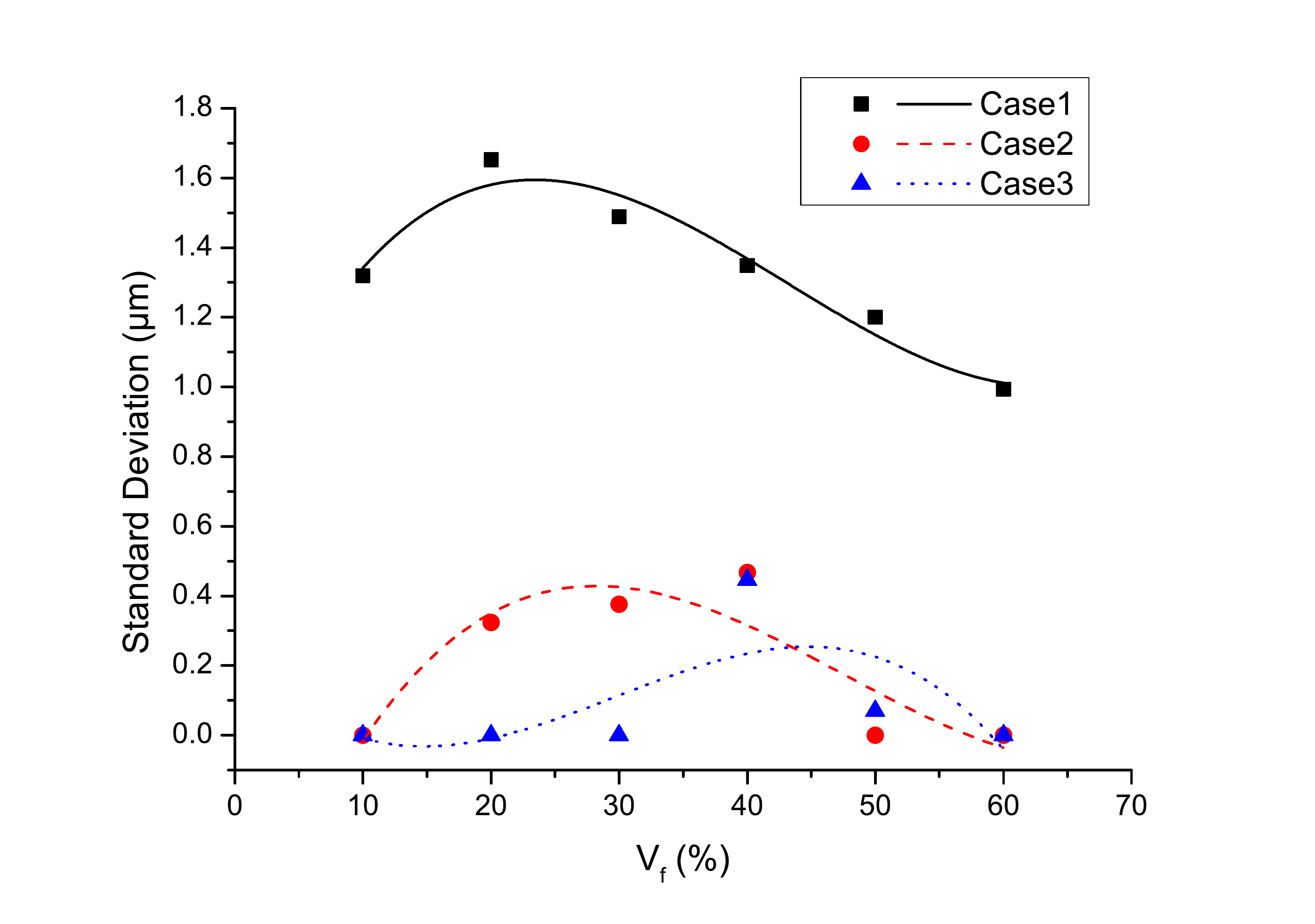}
		\label{fig:fig6b}
	}
	
	\caption{Variation of the Standard Deviation of (a) STC distribution for all three cases, (b) ineffective length with volume fraction in all three cases}
	
\end{figure}

\begin{figure*} 
	\centering
	\subfigure[]{\includegraphics[width=2in]{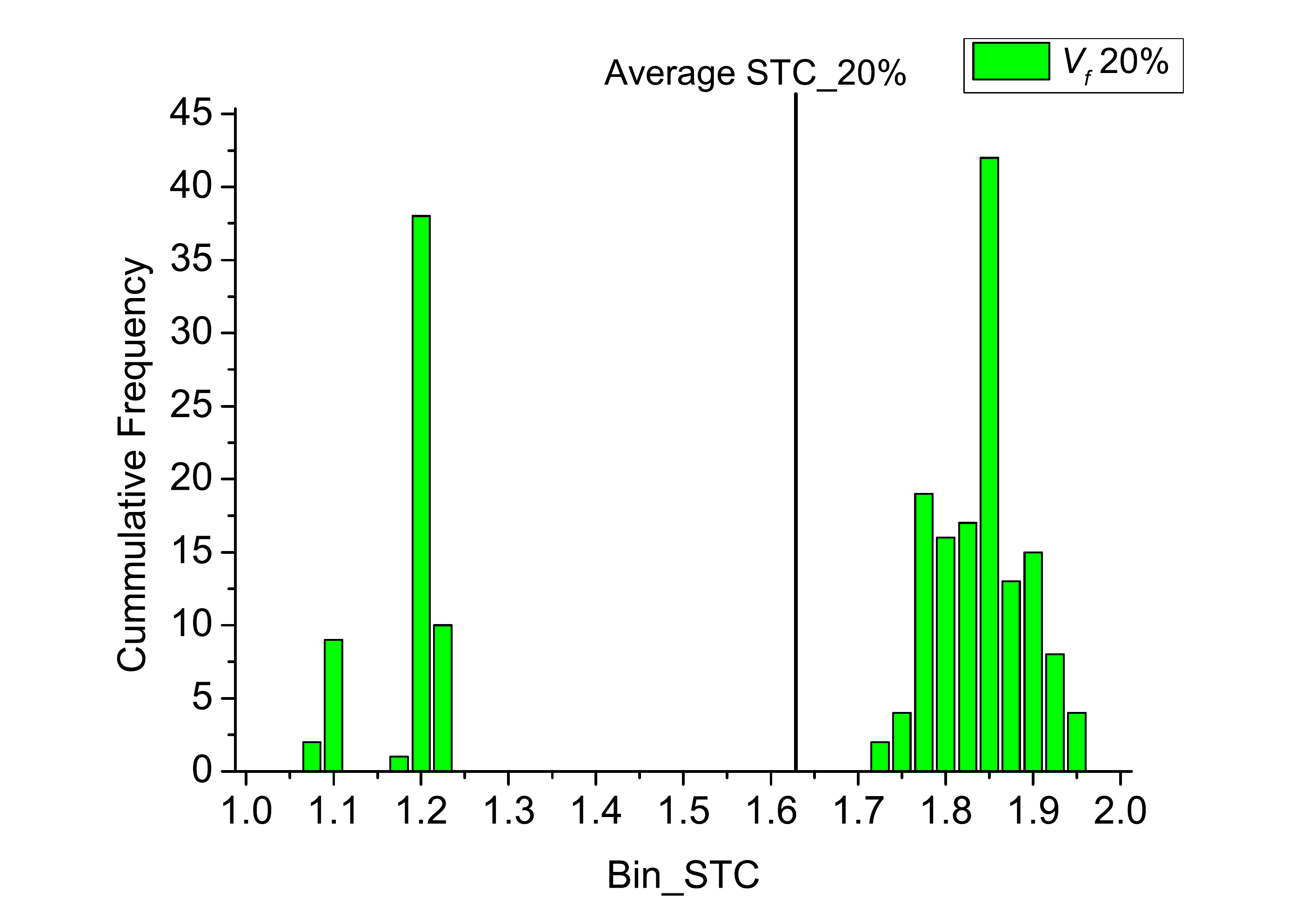}
	}
	~
	\subfigure[]{\includegraphics[width=2in]{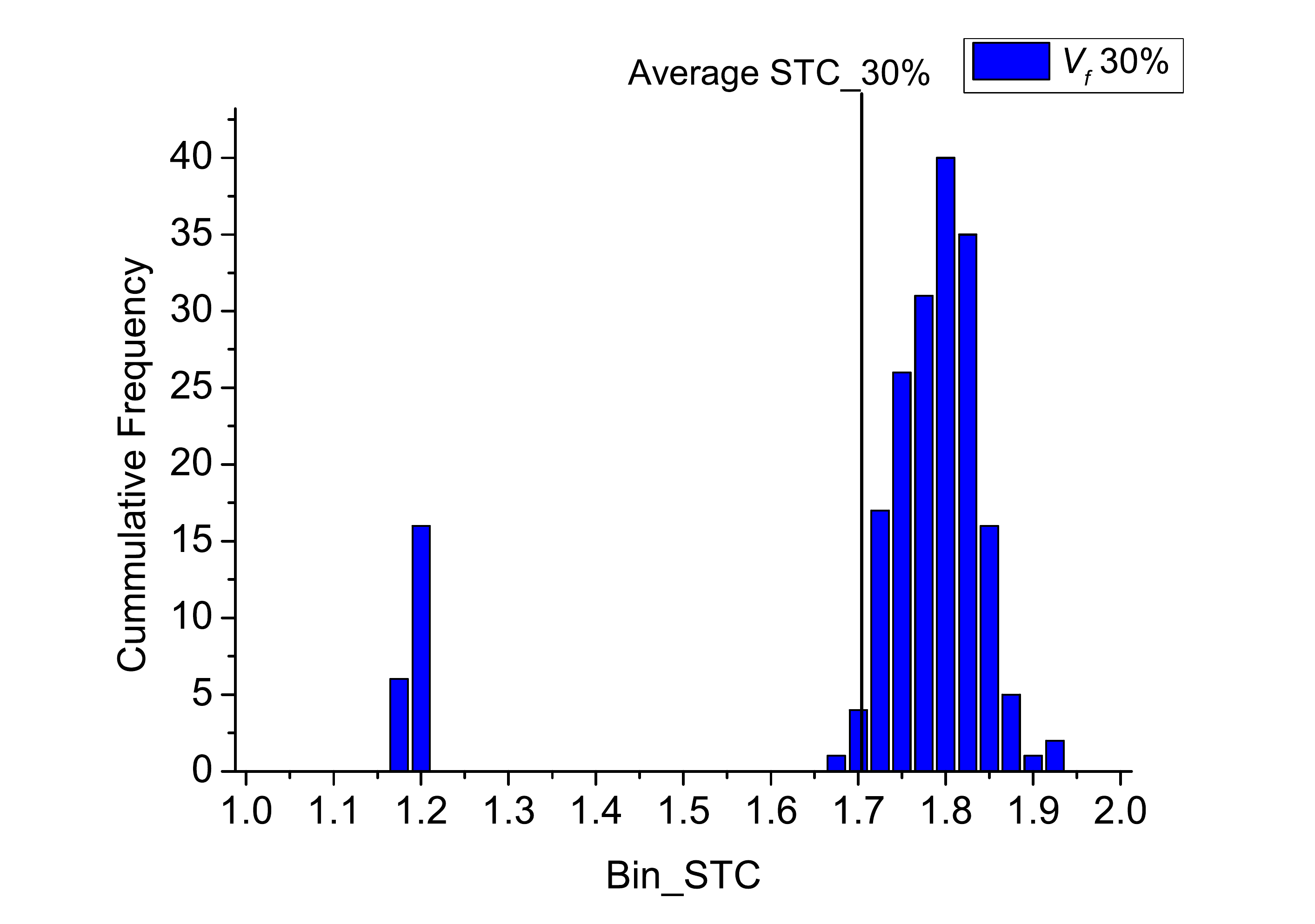}
	}
	~
	\subfigure[]{\includegraphics[width=2in]{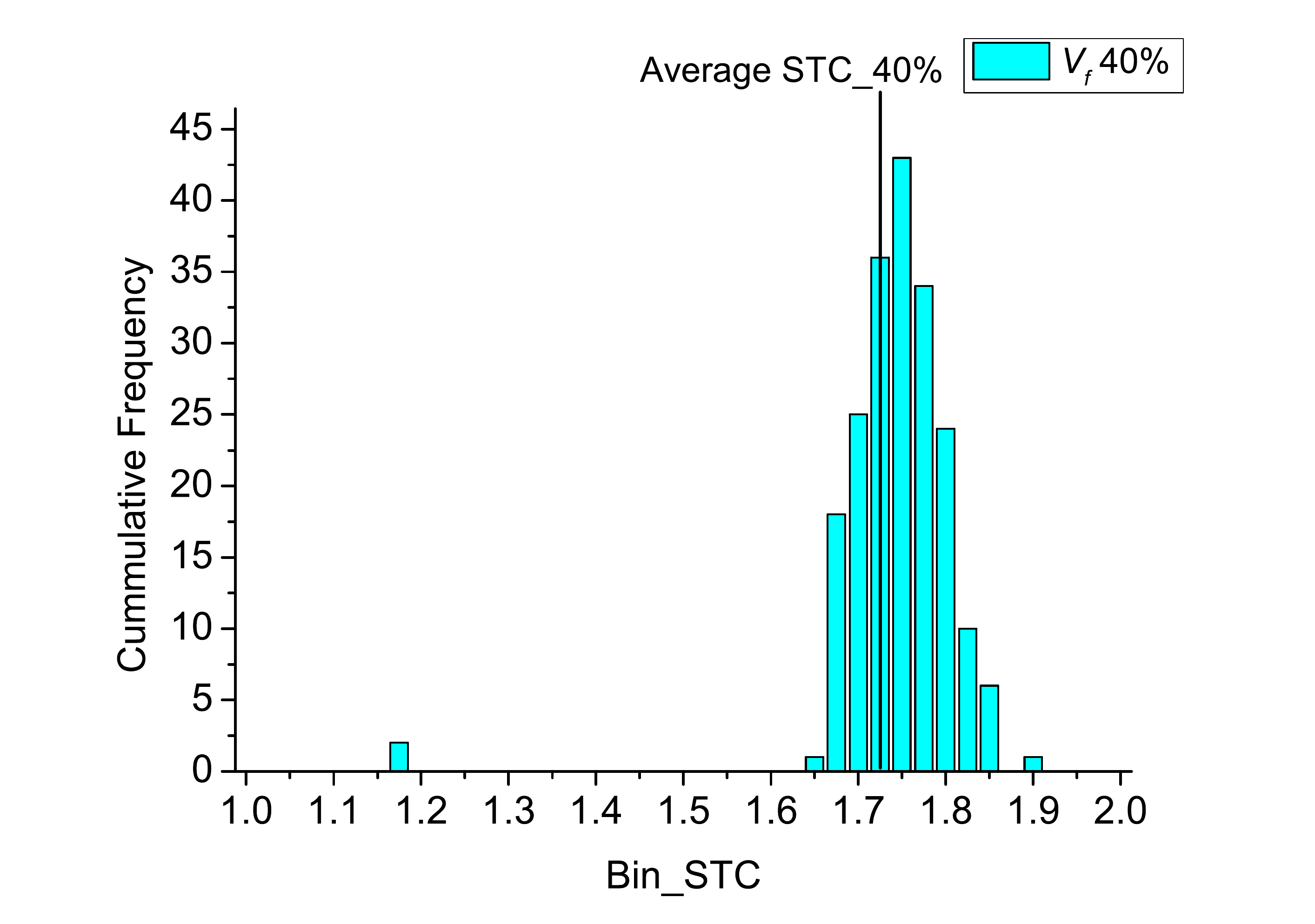}
	}
	~
	\subfigure[]{\includegraphics[width=2in]{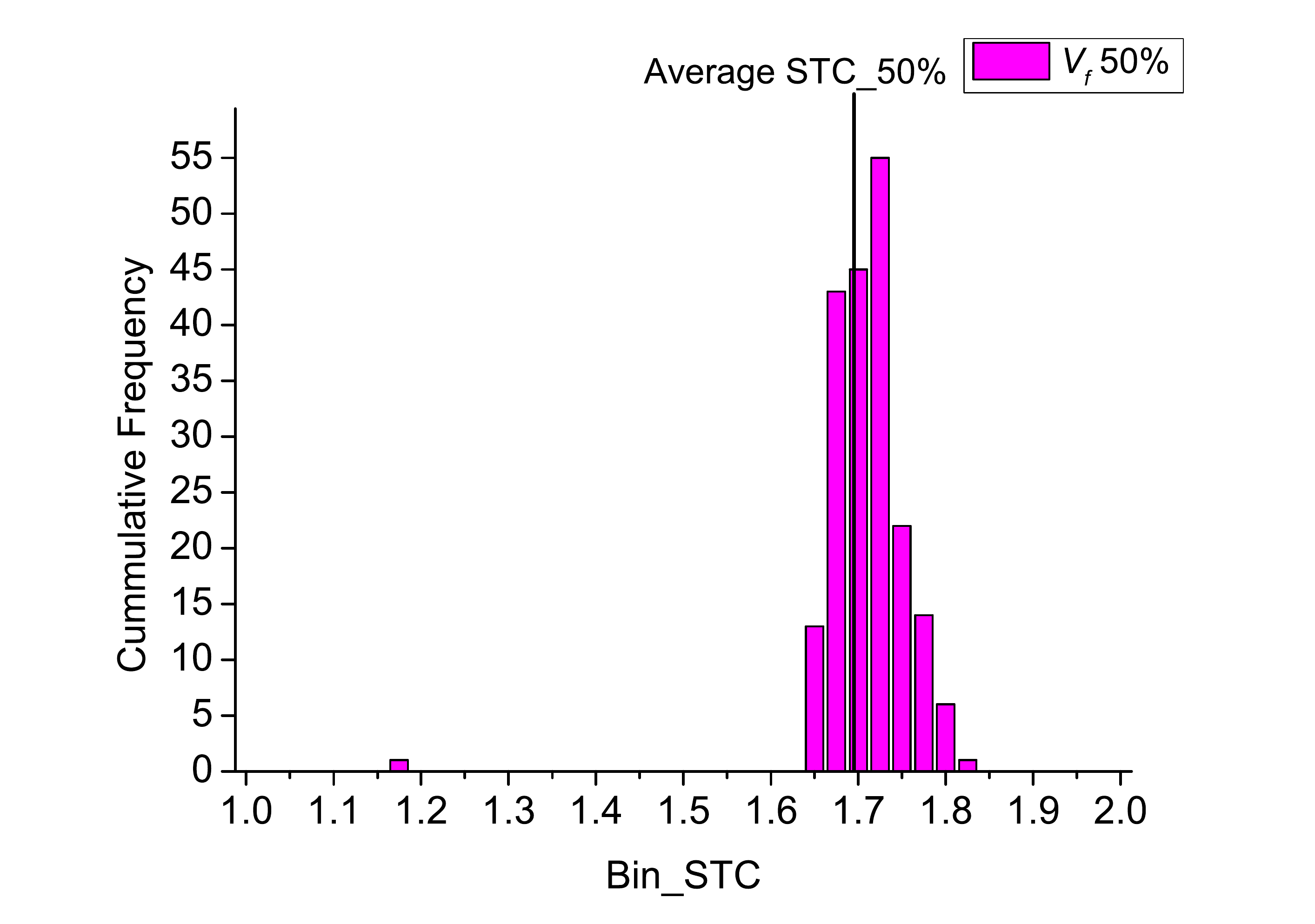}
	}
	~
	\subfigure[]{\includegraphics[width=2in]{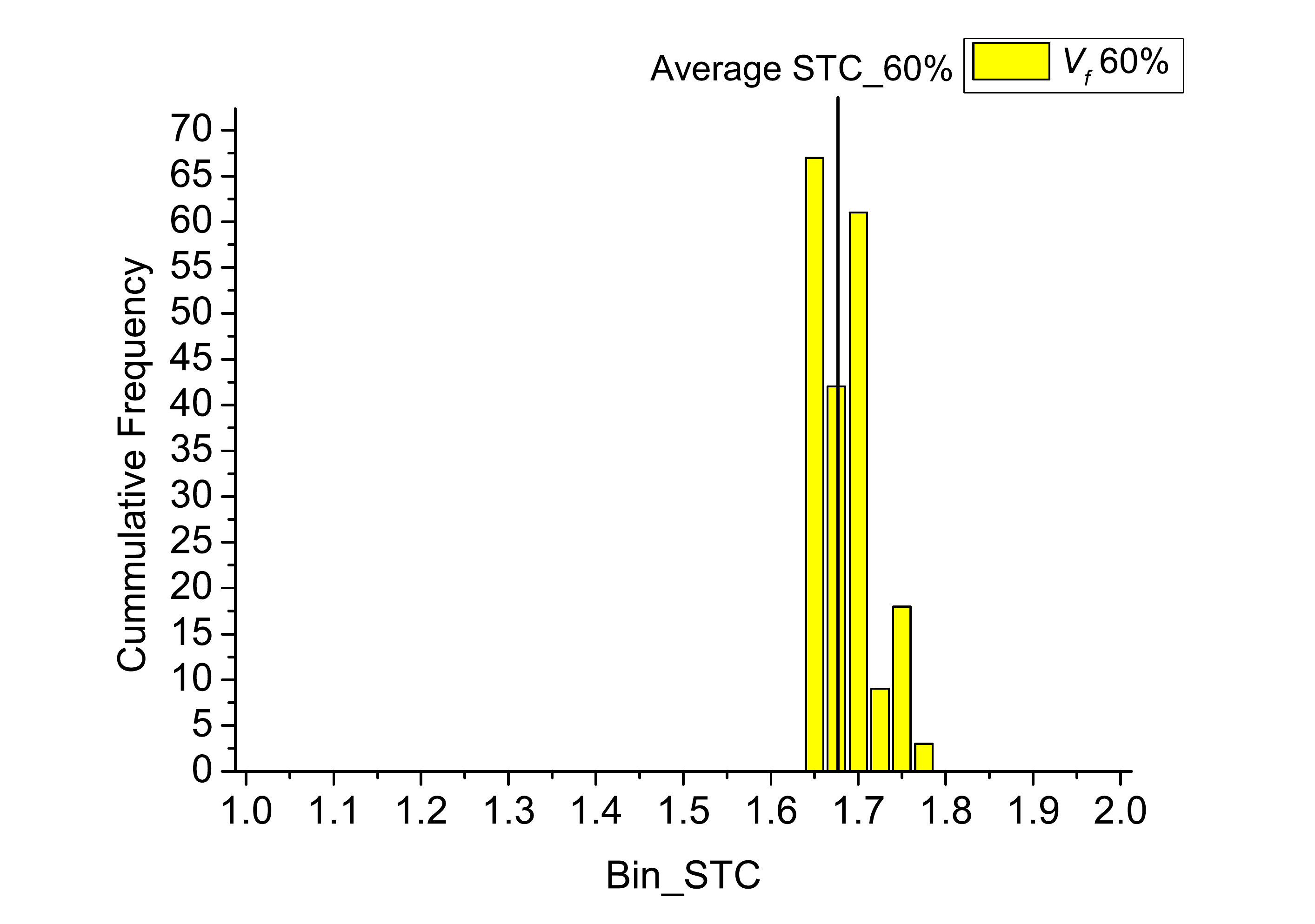}
	}
	~
	\subfigure[]{\includegraphics[width=2in]{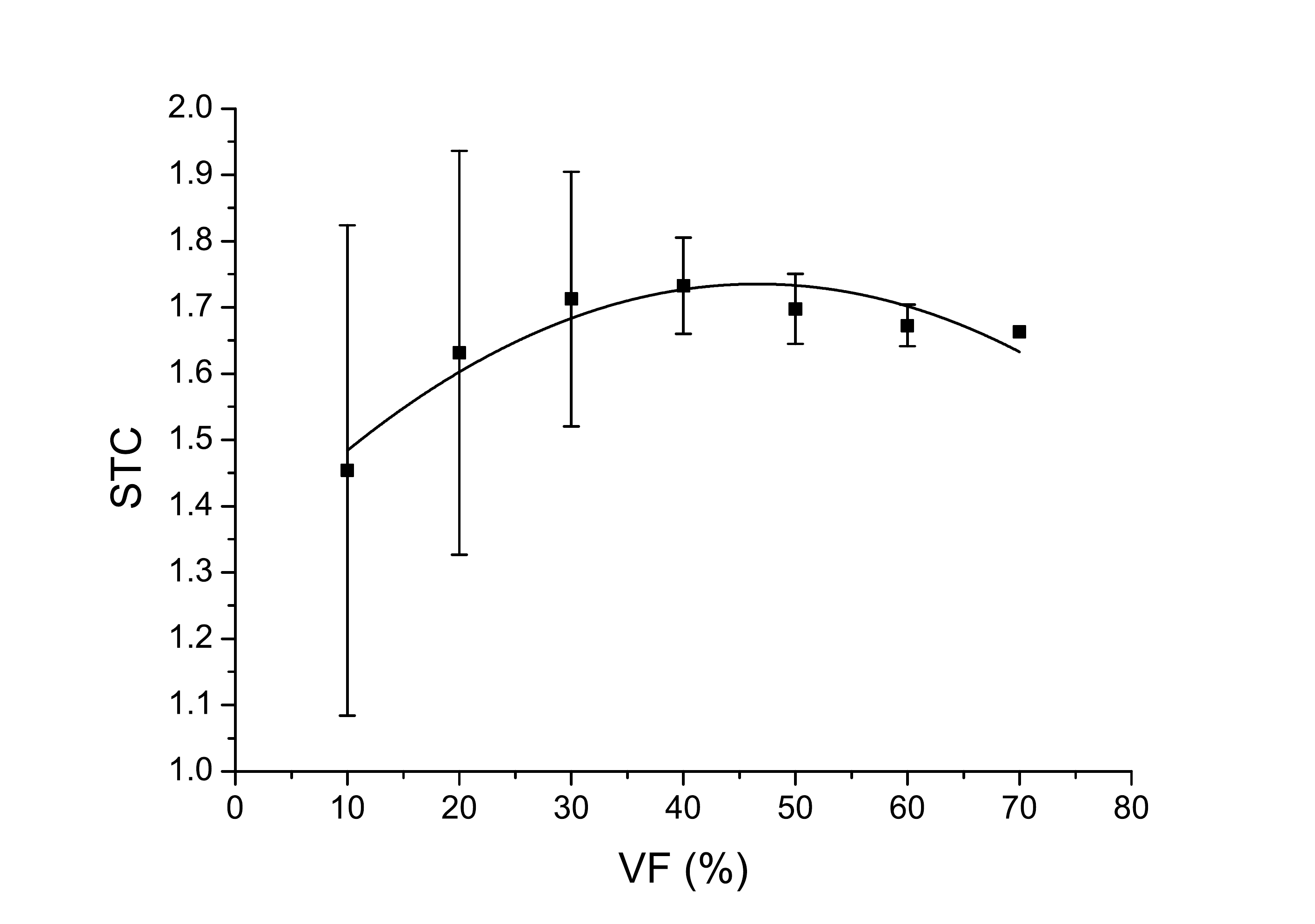}
	}
	\caption{Histogram plot of STC distribution for Case1 (a) $V_f$ $20\%$, (b)  $V_f$ $30\%$, (c)  $V_f$ $40\%$, (d)  $V_f$ $50\%$, (e)  $V_f$ $60\%$, (f)  Variation of average STC along with the standard deviation with respect to volume fraction.}
	\label{fig:fig7}
\end{figure*}

As discussed earlier the volume fraction, in this case, is varied ($V_f~10\%$ to $V_f~60\%$) by changing the number of fibers keeping the radius and the size of the RVE constant. The standard deviation of the STC values of the nearest fibers in 200 RVE’s is calculated and plotted against the corresponding fiber volume fraction as shown in Fig~\ref{fig:fig6a}. It is observed that the standard deviation of the STC distribution decreases with the increase in fiber volume fraction. To understand this behaviour, a complete set of STC data for each volume fraction is plotted in the form of the histogram, as shown in Fig~\ref{fig:fig7}, and is analysed. Looking at the histograms in Fig~\ref{fig:fig7}, it is observed that, there are 3 bin ranges where the cumulative frequency of the STC has the values viz. $1.075-1.1,~ 1.175-1.225$ and $1.65-1.95$. Table 3 shows the frequency of STC falling in these ranges for each volume fraction. The values in bin range 1 and 2 decrease with the fiber volume fraction but the values in bin range 3 increase with volume fraction.


\begin{figure}
	\centering
	\includegraphics[width=3 in]{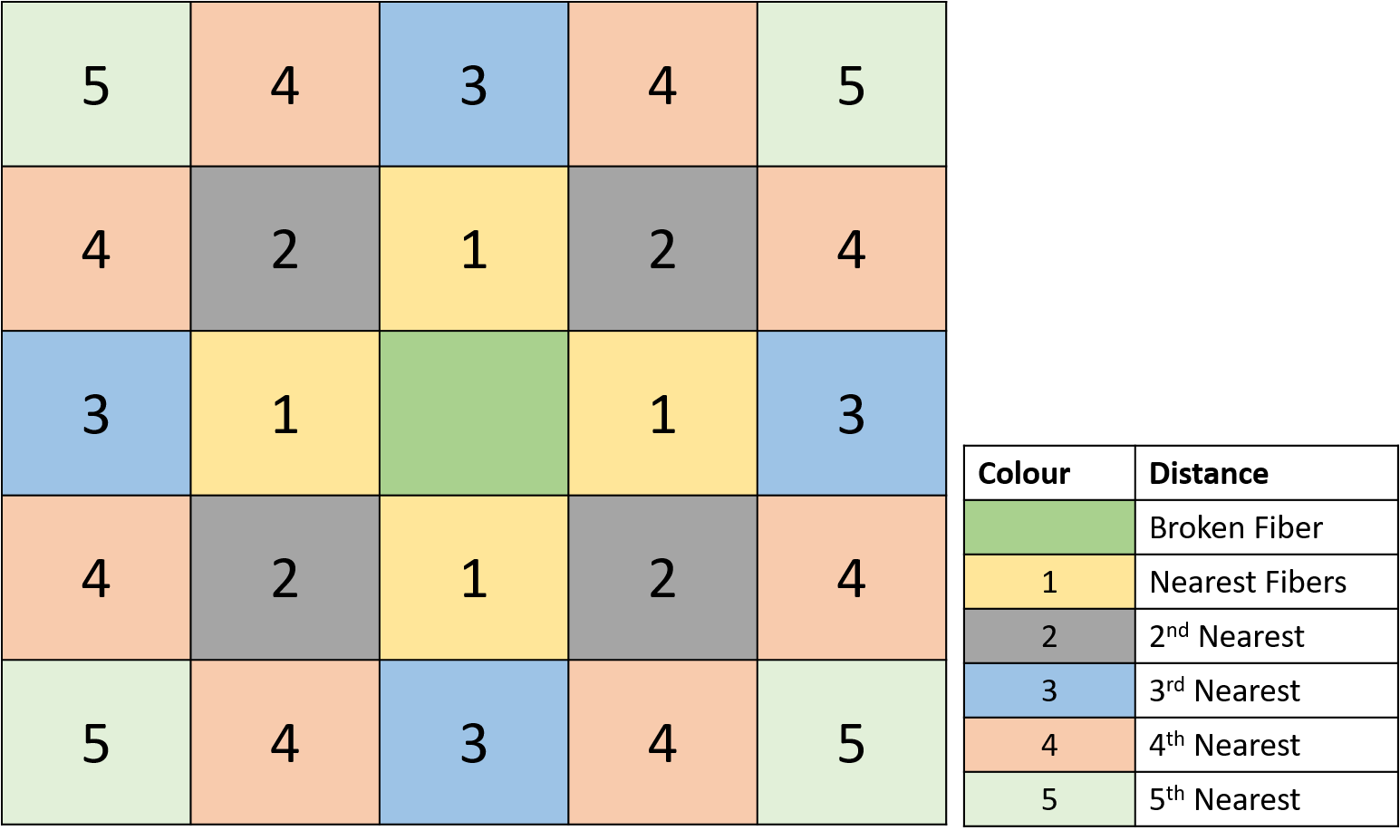}
	\caption{Representation of RVE with 5 locations marked by numbers 1 to 5 based on the centre distance of fibers from the broken fiber.}
	\label{fig:fig8}
\end{figure}

To get the sense of the distribution of STC for a volume fraction due to the random arrangements of fibers, a brief overlook of the distribution of fibers in an RVE is discussed. In Fig~\ref{fig:fig8} there are five locations named as $1-5$ depending on their normal distance from the broken fiber and also they are colour coded. These five locations accommodate the fibers. The stress transfer from the broken fiber decreases from location-1 to location-5. If the fiber volume fraction of $70~ \%$ is considered, the RVE must have 25 number of fiber including the central broken fiber, as shown in Table~\ref{tab:tab1}. Now, in this fully populated RVE we can name the fibers as $1^{st}$ nearest fibers, $2^{nd}$ nearest, and so on up to $5^{th}$ nearest depending on their distance from the broken fiber. In the fully populated RVE ($V_f~70\%$) the $1^{st}$ nearest fiber, $2^{nd}$ nearest, $3^{rd}$ nearest, etc. occupy location 1, 2, 3, 4, and 5 respectively, but the same is not true for the RVE's with lower volume fractions i.e. the $1^{st}$ nearest fiber may or may not be in location 1 and similarly for other nearest fibers. Thus we can say that for any fiber except broken fiber there are these 5 locations to be placed randomly.

\begin{table}
	\centering
	\caption{Distribution of STC values for Case 1 in 3 bin ranges}
	\label{tab:tab5}       
	\begin{tabular}{l c c c c }
		\hline\noalign{\smallskip}
		\multirow{2}{*}{$V_f$ (\%)} & \multicolumn{3}{c}{Bin Range} & \multirow{2}{*}{Total} \\
		& 1.075-1.1&1.175-1.225&1.65-1.95& \\
		\hline\noalign{\smallskip}
		10&67&47&86&200 \\
		20&11&49&140&200\\
		30&0&22&178&200\\
		40&0&2&198&200\\
		50&0&1&199&200\\
		60&0&0&200&200\\
						
		\noalign{\smallskip}\hline
	\end{tabular}
\end{table}

Now, coming back to the distribution of STC for a volume fraction, the RVE’s shows a particular pattern for each bin range. For range 1 which is seen in volume fraction $10$ and $20~\%$, it is observed that the RVE’s corresponding to this range does not have fibers at the location 1 and 2, but are present either at locations 3, 4 and/or 5. Thus in these RVE’s the distance of the $1^{st}$ nearest fiber is more which results in a low STC value. The RVE’s corresponding to bin range 2 does not have fibers at location 1 but, all other locations are occupied. In these RVE’s the nearest fiber distance is less as compared to bin range 1 RVE’s, hence the STC value is more. Bin range 3 has the highest STC values because there is at least one fiber at location 1 which is at the minimum distance from the broken fiber in all the RVE’s corresponding to it.

The standard deviation of STC distribution for volume fraction $10~\%$ is high; as in this volume fraction the number of fibers are less and as there are in all 24 locations excluding the broken fibers to be filled the number of ways they can be arranged at different locations are more. As a result, all three bin ranges have values that imply, that there is a lot of variation in STC values and hence we get the highest value of standard deviation. As the volume fraction is increasing the number of values in range 1 and range 2 can be seen to be decreasing and hence the variation in the value of STC is less resulting in low values of standard deviation. In other words, we can say that with the increase in fiber volume fraction the RVE is tending towards periodicity.

\begin{figure*} 
	\centering
	\subfigure[]{\includegraphics[width=0.31\textwidth]{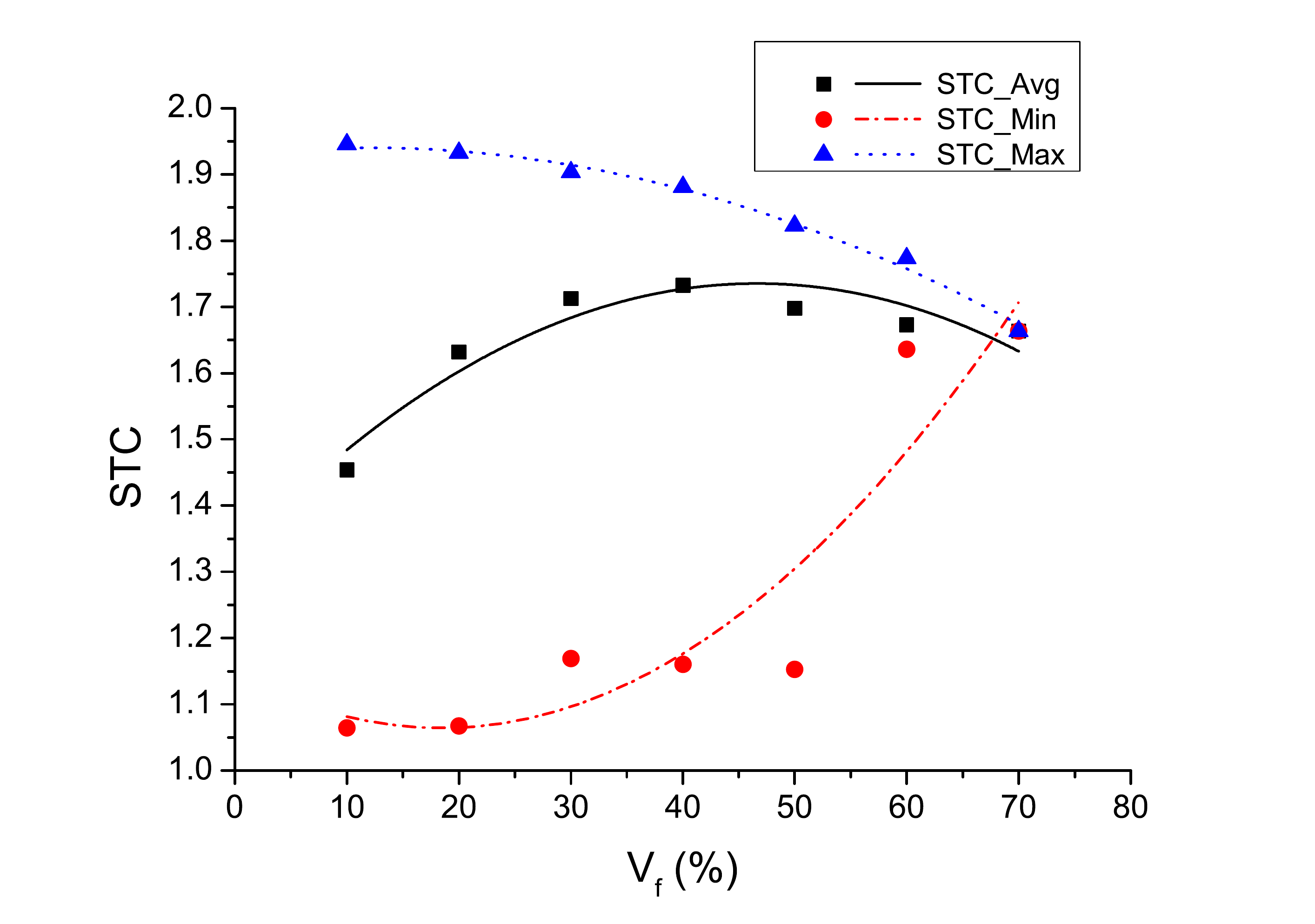}
		\label{fig:fig9a}
	}
	~
	\subfigure[]{\includegraphics[width=0.31\textwidth]{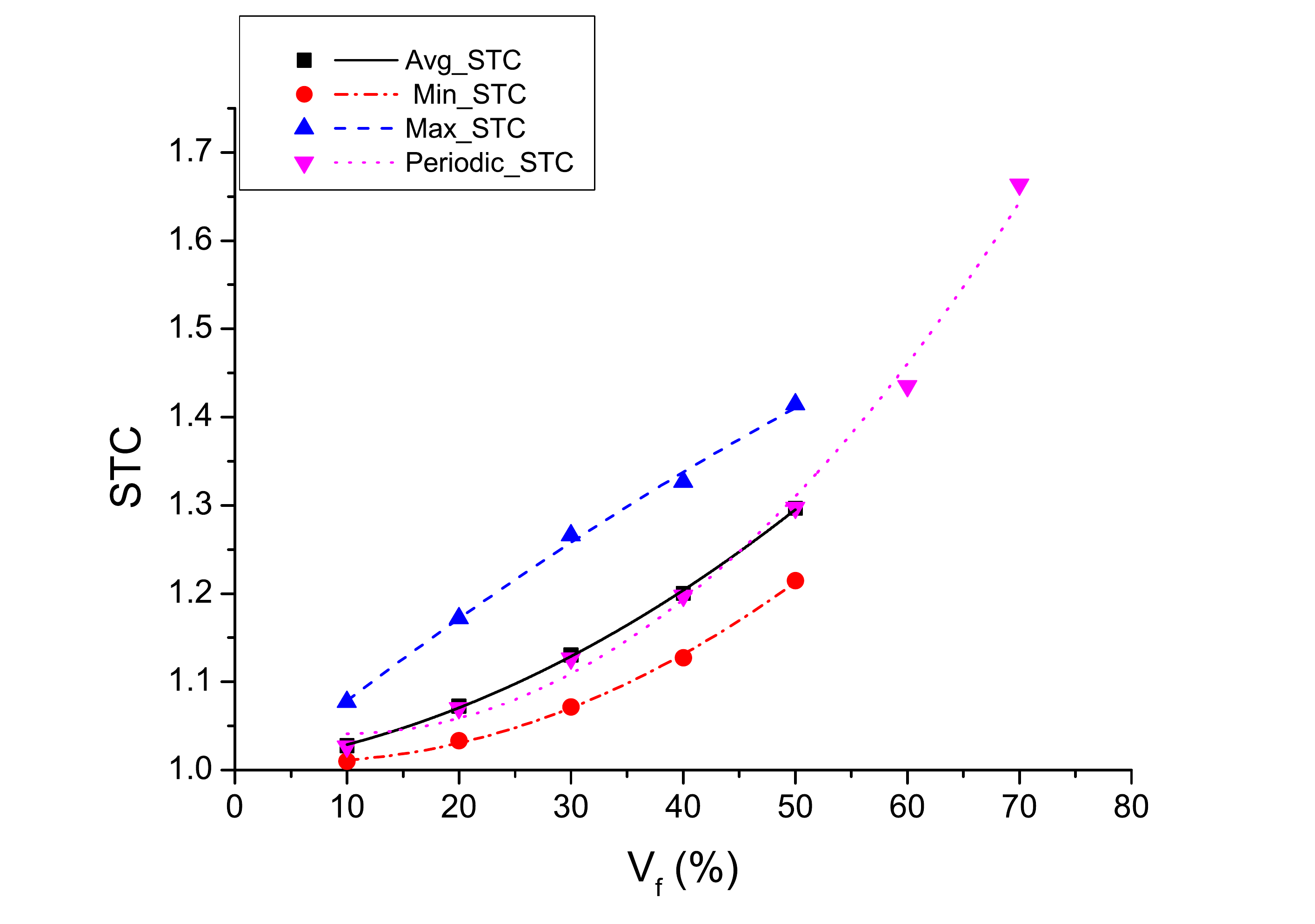}
		\label{fig:fig9b}
	}
	~
	\subfigure[]{\includegraphics[width=0.31\textwidth]{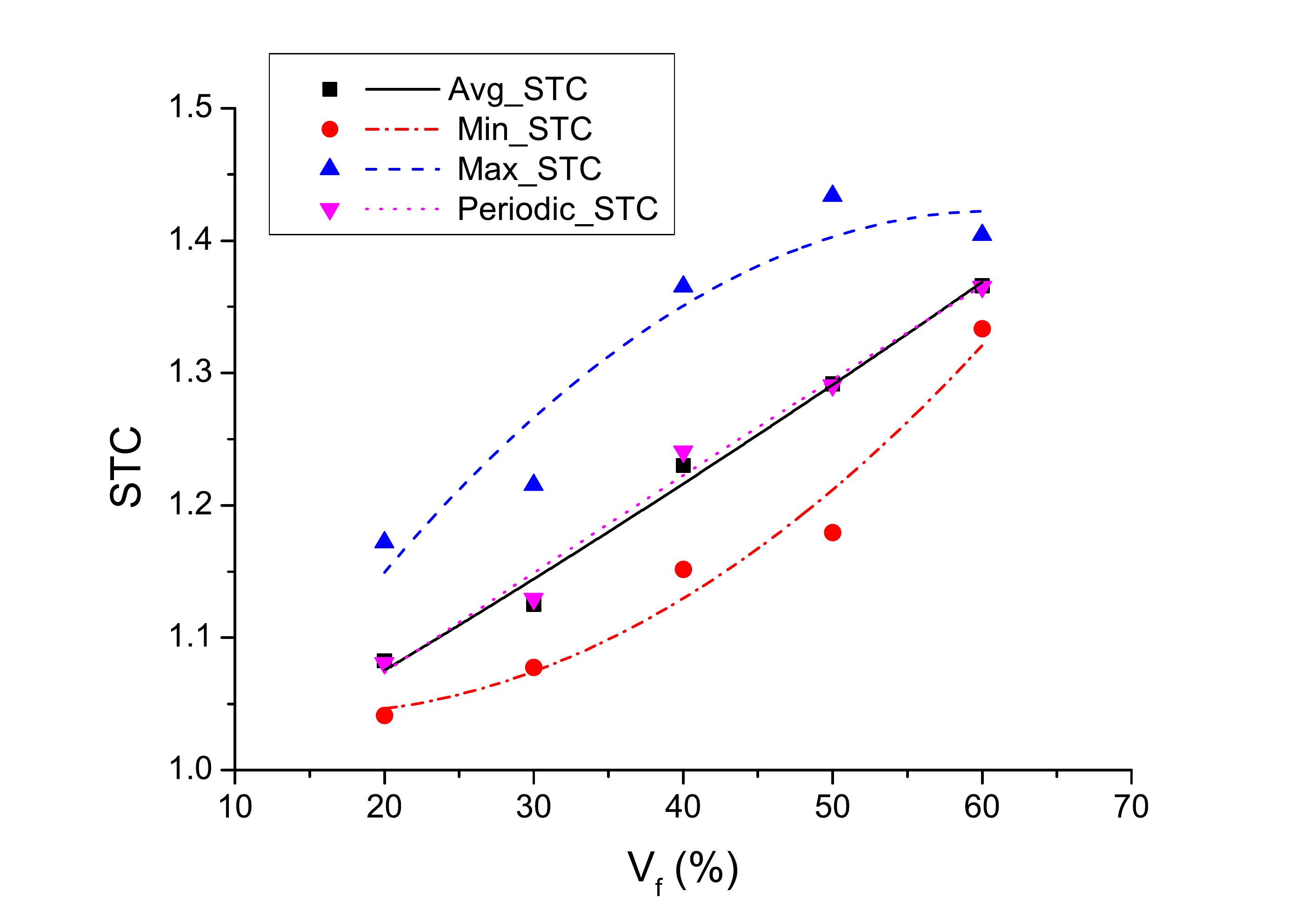}
	\label{fig:fig9c}
	}
	
	\caption{Variation of the average, minimum and maximum STC with volume fraction. Also for Case 2 and Case 3 variation of STC for periodic distribution is plotted. (a) Case 1, (b) Case 2, (c) Case 3}
	
\end{figure*}

Analysing the distribution of STC, now the effect of volume fraction on the average value of STC can be predicted. Fig~\ref{fig:fig9a} shows the variation of average, maximum and Minimum STC against the fiber volume fraction. It is observed that the average STC first increases with the volume fraction up to $40~\%$ and then it decreases whereas the maximum STC decreases monotonously and minimum STC increases. The minimum values of STC in fiber volume fraction $10~\%$ as stated above is because the nearest fiber is not at location 1 but is at location 4 or 5. Since the distance between the nearest fiber from the broken fiber is very large, the stress transferred from the broken fiber is very less; as the maximum part of it is depleted in the shearing of the matrix. The maximum value of STC is observed when the nearest fiber is present at location 1. As the volume fraction is increasing because the number of fibers is increasing, the probabilities of the fibers falling at location 1 are increasing. Hence the minimum values are increasing. However, a reverse trend for maximum value is observed i.e. their values increases with volume fraction. The reason to this opposite trend is the shielding effect~\cite{Swolfs2013}. 

Again consider the case of volume fraction $10~\%$, if the $1^{st}$ nearest fiber is at location 1 it will result in maximum stress but only when there is only one fiber at this location and all other positions corresponding to location 1 are empty. As the number of $1^{st}$ nearest fiber increases, the stress transferred from the broken fiber is shared with all of them and hence the STC will be less, but, still the value will be high to fall in bin range 3. This decrease in the transferred value of stress due to more number of fibers surrounding the broken fiber is called the shielding effect. $2^{nd}$ nearest fibers also contribute to this effect. The contribution of $3^{rd}$,$4^{th}$ and $5^{th}$ is less. The shielding effect is the reason for having variation in the values in all bin ranges especially in range 3.

Another parameter that is affected along with the STC due to the distribution of the fibers and the volume fraction is the ineffective length. When a fiber breaks in the microstructure of the composite, its strength (load carrying capacity) in the loading direction reduces to zero. But as the loading continues the fiber starts regaining its strength. This regaining of strength is because of the shearing of the matrix. It is assumed that the bond between matrix and fiber is perfect. The surrounding fibers play a very important role in the fast gaining of strength. For better understanding, three scenarios are discussed below. The boundary conditions are the same as mentioned in section~\ref{sec:matprop}.
\paragraph{Number of neighbouring fibers $= 0$}:
In this case, there is only one fiber in the RVE which is broken. As the loading continues the fiber tries to move out of the RVE since there is no restriction at the $z= 0$ plane to restrict the applied displacement. The tendency of the fiber to displace will stretch the matrix due to the perfect bond between them. The stretching of the matrix imposes restrictions on the fiber from displacing as at $z=0$ plane the matrix has zero degrees of freedom in the loading direction. Due to restriction, the fiber starts building the stress in the loading direction. In other words, the fiber starts regaining its strength to oppose the applied displacement. In this case, the rebuilding of the fiber strength is all because of the matrix.
\paragraph{Number of neighbouring fibers $= 1$}:
The non-broken neighbouring fiber present inside the region of the shearing of the matrix will act as a catalyst in building the stress in the broken fiber. This is because the intact fiber resists the shearing of the matrix due to the perfect bond between them along the length. Thus this fiber will give an extra hand to the matrix in holding and building the stress in the broken fiber.
\paragraph{Number of neighbouring Fibers $> 1$}:\label{para:c}
The more the number of non-broken neighbouring fibers present within the shielding region of the matrix the less shearing of the matrix will suffice the recovering of the strain in the broken fiber.\\
 
The length of the broken fiber which is responsible for the shearing of the matrix before it regains its strength to oppose the applied load is, in fact, the length of the broken fiber which does not support any loading, and hence it is called ineffective length. Thus, in case with no neighbouring fibers, ineffective length will be very large and scenario with more than one neighbouring fibers will show the least ineffective length, the ineffective length with only one neighbouring fiber lies in between.

The values of ineffective length for 200 RVE’s for each volume fraction is obtained with the help of an algorithm written in the form of code in \textit{Python}. This algorithm obtains the values of normal stresses, $S_{33}$, in the longitudinal direction of the fiber on the broken fiber at an interval of $1 ~\mu m$. After this, the stress at the far-field i.e. at $z=100 \mu m$ is obtained and as per the definition, it is multiplied with 0.9. The value obtained after multiplication is matched with the stress values obtained along the length. The nearest value of the stress close to the multiplied value is obtained and the corresponding length of the fiber is the ineffective length.\footnote{The least value of the length measured is $1 \mu m$, so the ineffective value is the rounded value corresponding to the multiplied value of the stress.}

The standard deviation of ineffective lengths of all 200 RVE’s in a particular volume fraction is plotted against the fiber volume fraction as shown in Fig~\ref{fig:fig6b}. Depending on the locations of the neighbouring fibers in the RVE as 1-5 referring to Fig.~\ref{fig:fig8}, the value of ineffective length changes similar to STC. In fiber volume fraction of $10\%$, as there are only 3 fibers excluding the broken fiber to be placed in the RVE, there are numerous ways this can be done. The smallest value is observed when all three fibers are present at location 1. The next smallest value is observed in the RVE’s where there are two fibers at location 1 and the third fiber is present at any locations 2-5. The next smallest values are observed in the RVE’s in which there were only one fiber at location 1 and the other two fibers at any other locations. The location of these two fibers is responsible for the large variation seen in this range of ineffective length. RVE’s with no fiber at location 1 results in the 2nd largest value but at least one fiber is found to be present at location 2. The largest value of ineffective length is observed when there is no fiber present at locations 1 and 2 but are present at locations 3, 4, or 5.

The standard deviation of the ineffective length is highest for the fiber volume fraction of $20\%$. The reason for this is there are 6 fibers to be placed excluding the broken fibers. Unlike for $10\%$ volume fraction, there can be four fibers taking location 1 in the RVE resulting in the lowest value of the ineffective length. As the number of fiber at location 1 is reducing from 4 to 0 the RVE’s will have higher values of ineffective length respectively. Also in this volume fraction the number of fibers at location 2, if no fiber is at location 1, will affect the value of the ineffective length. The largest value of ineffective length is observed when there are no fibers present at locations 1 and 2. Since there is a lot of variation in the position of the fibers at locations 1 and 2 this volume fraction shows the highest standard deviation value.

\begin{figure*} 
	\centering
	\subfigure[]{\includegraphics[width=0.31\textwidth]{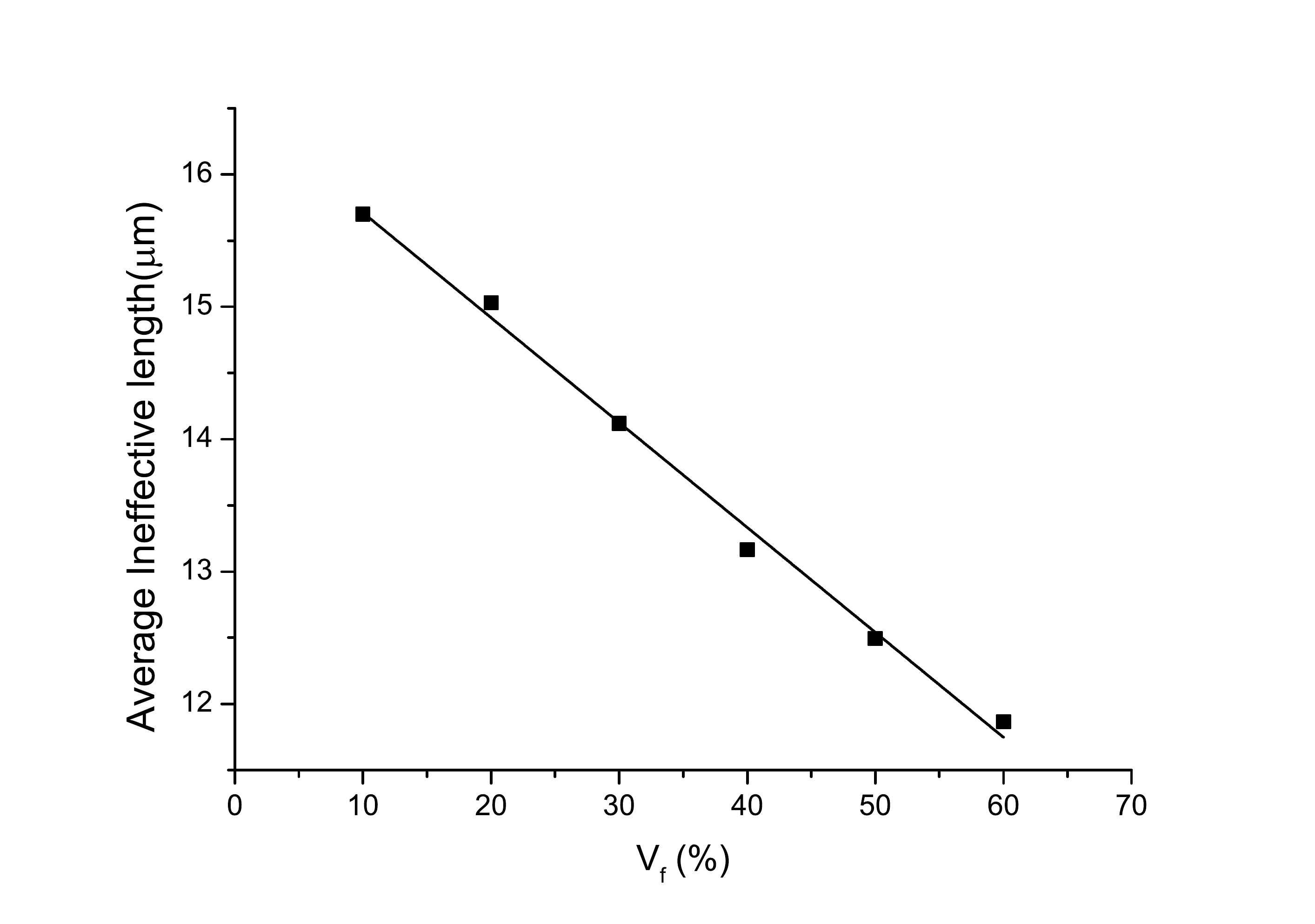}
		\label{fig:fig10a}
	}
	~
	\subfigure[]{\includegraphics[width=0.31\textwidth]{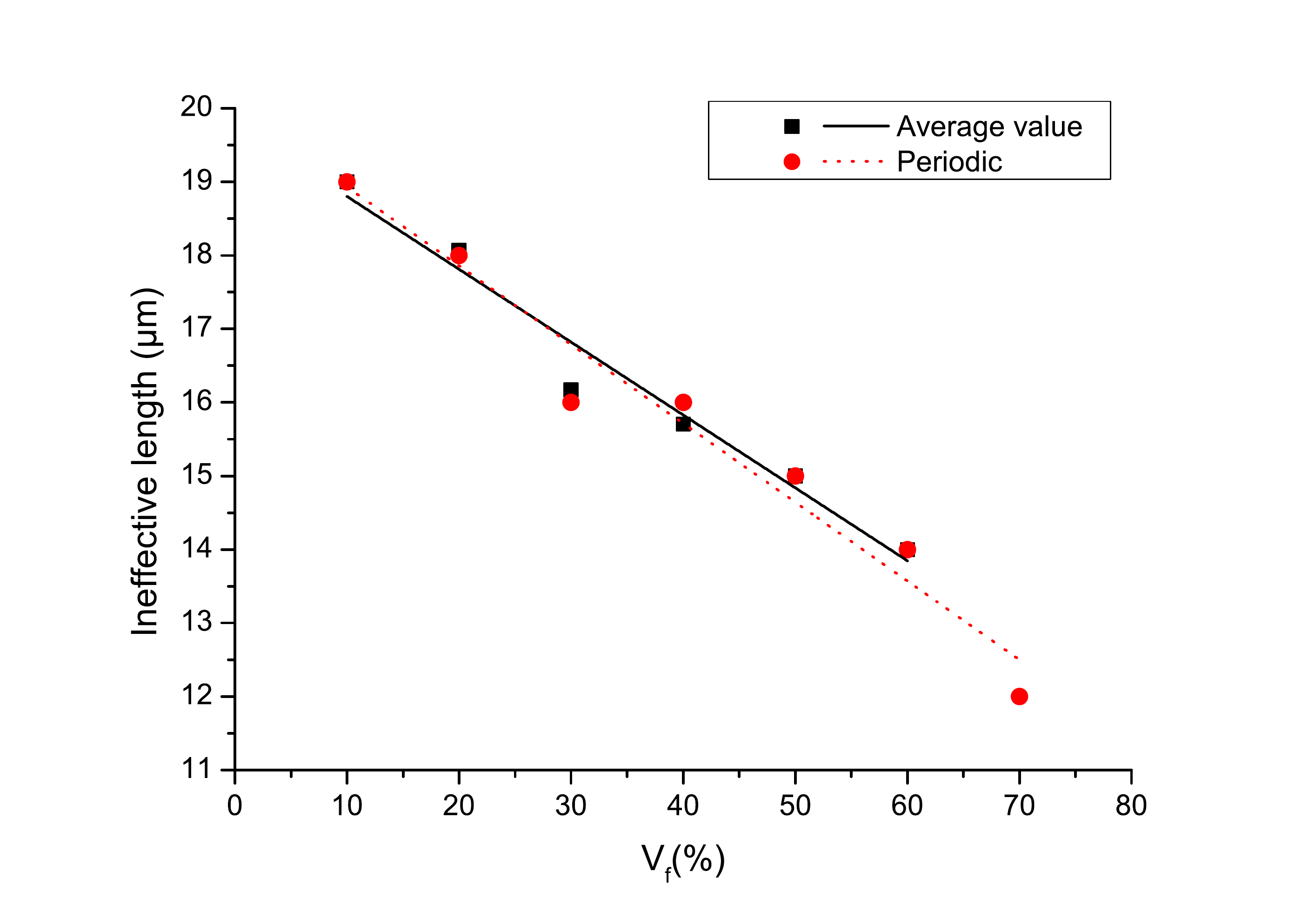}
		\label{fig:fig10b}
	}
	~
	\subfigure[]{\includegraphics[width=0.31\textwidth]{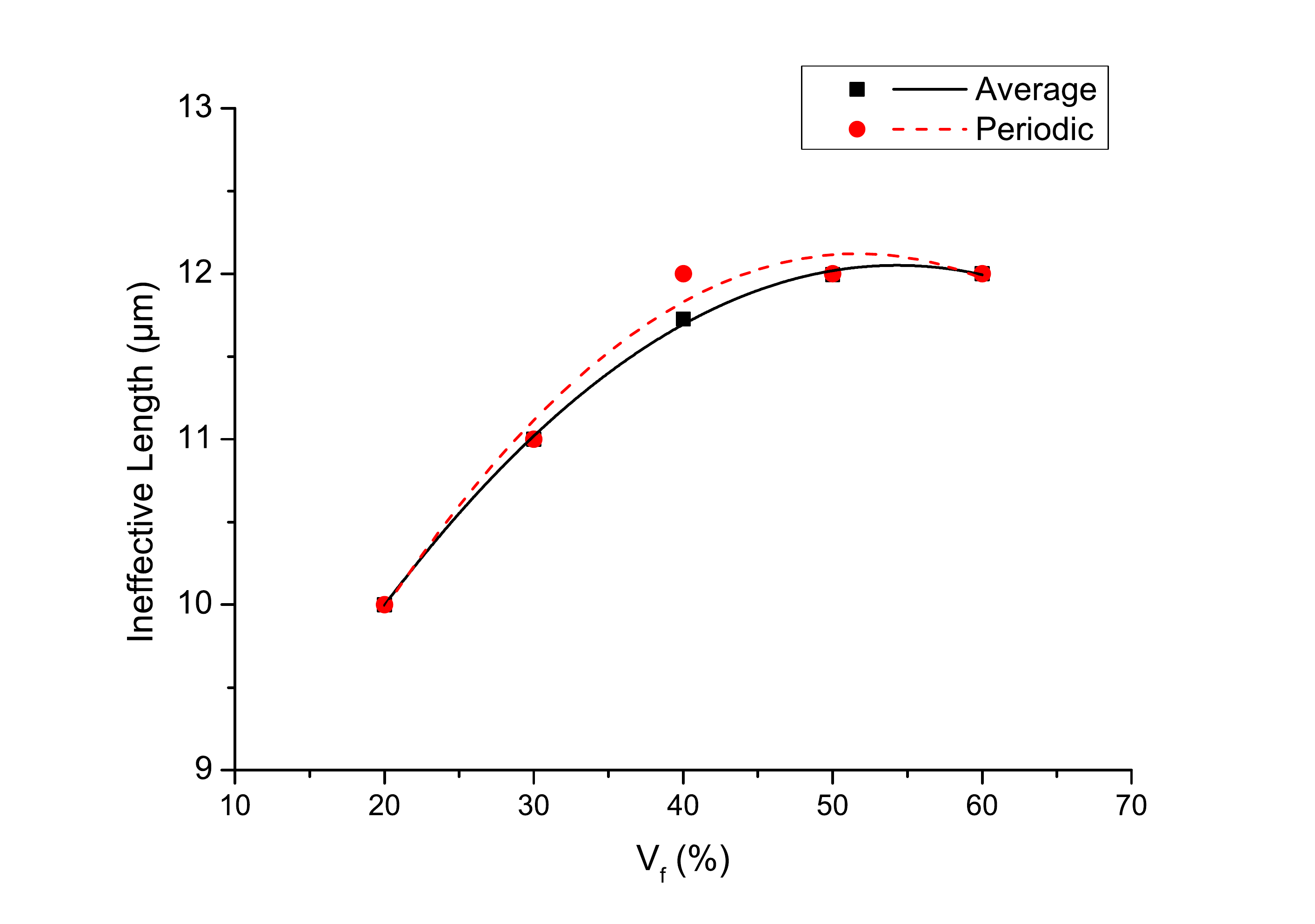}
		\label{fig:fig10c}
	}
	
	\caption{Variation of the average ineffective length with volome fraction. Also for Case 2 and Case 3 variation of ineffective length for periodic distribution is plotted. (a) Case 1, (b) Case 2, (c) Case 3}
	
\end{figure*}

As the volume fraction is further increased with the increase in the number of fibers, as shown in Table~\ref{tab:tab1} the probability of the fibers occupying locations 1 and 2 increases. This increase in the probability will result in smaller values of ineffective length. Also, the variation in the values will be less as the RVE will tend to become more periodic. It is observed that the frequency of the values corresponding to the largest ineffective length in fiber volume fraction $10\%$ is more than the frequency corresponding to the smallest value. But as the fiber volume fraction is increasing this trend is observed to get reversed. In fiber volume fraction $60\%$ there are more number of values corresponding to the smallest value and the frequency corresponding to the largest value is less. The average values of the ineffective length for 200 RVE’s in each volume fraction are plotted against the volume fraction, as shown in Fig.~\ref{fig:fig10a}. From this figure, it can be seen that, as the number of fibers increases around the broken fiber, the average ineffective length decreases. 

\subsection{Analysis of Case 2.}

\begin{figure*} 
	\centering
	\subfigure[]{\includegraphics[width=2in]{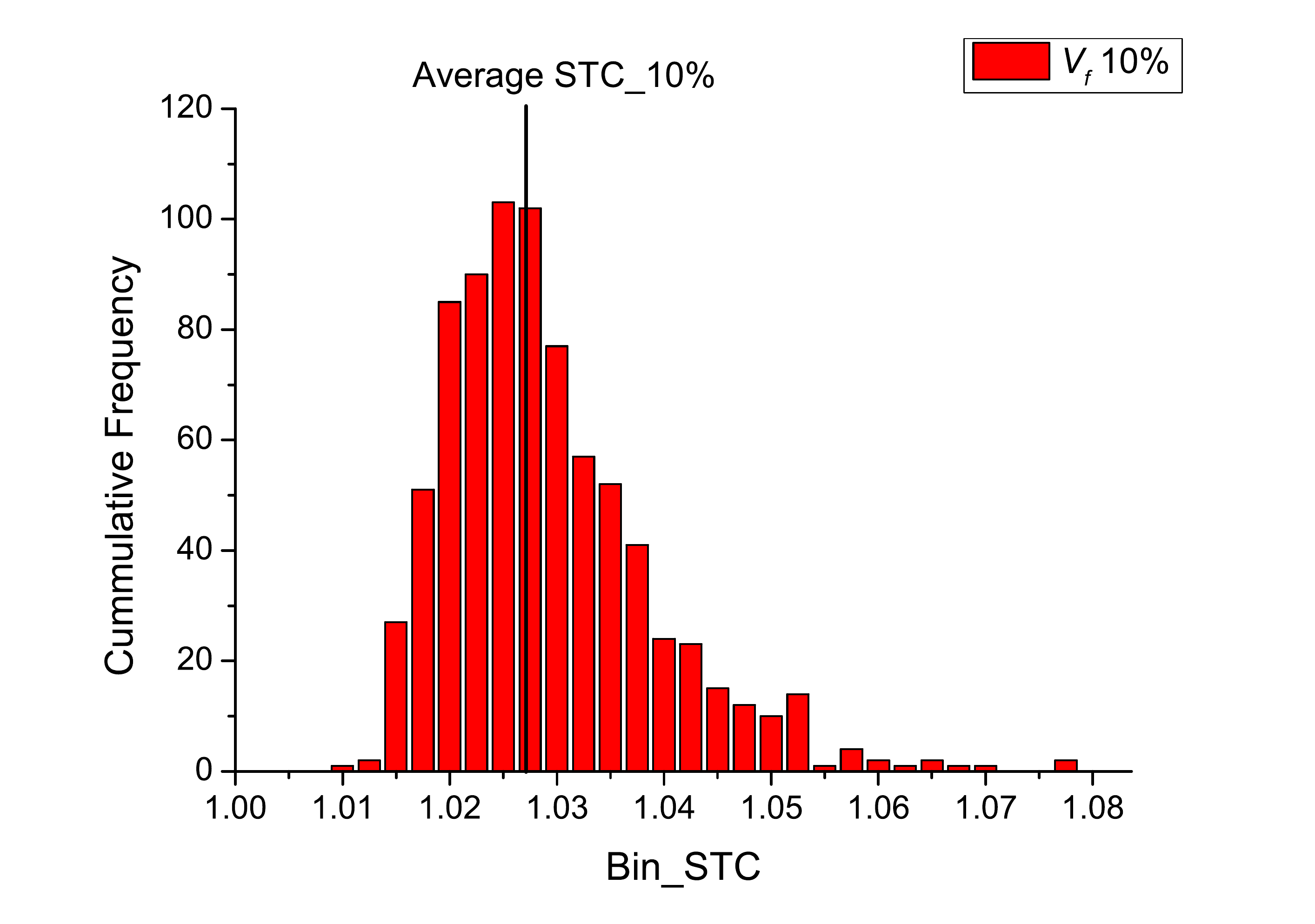}
	}
	~
	\subfigure[]{\includegraphics[width=2in]{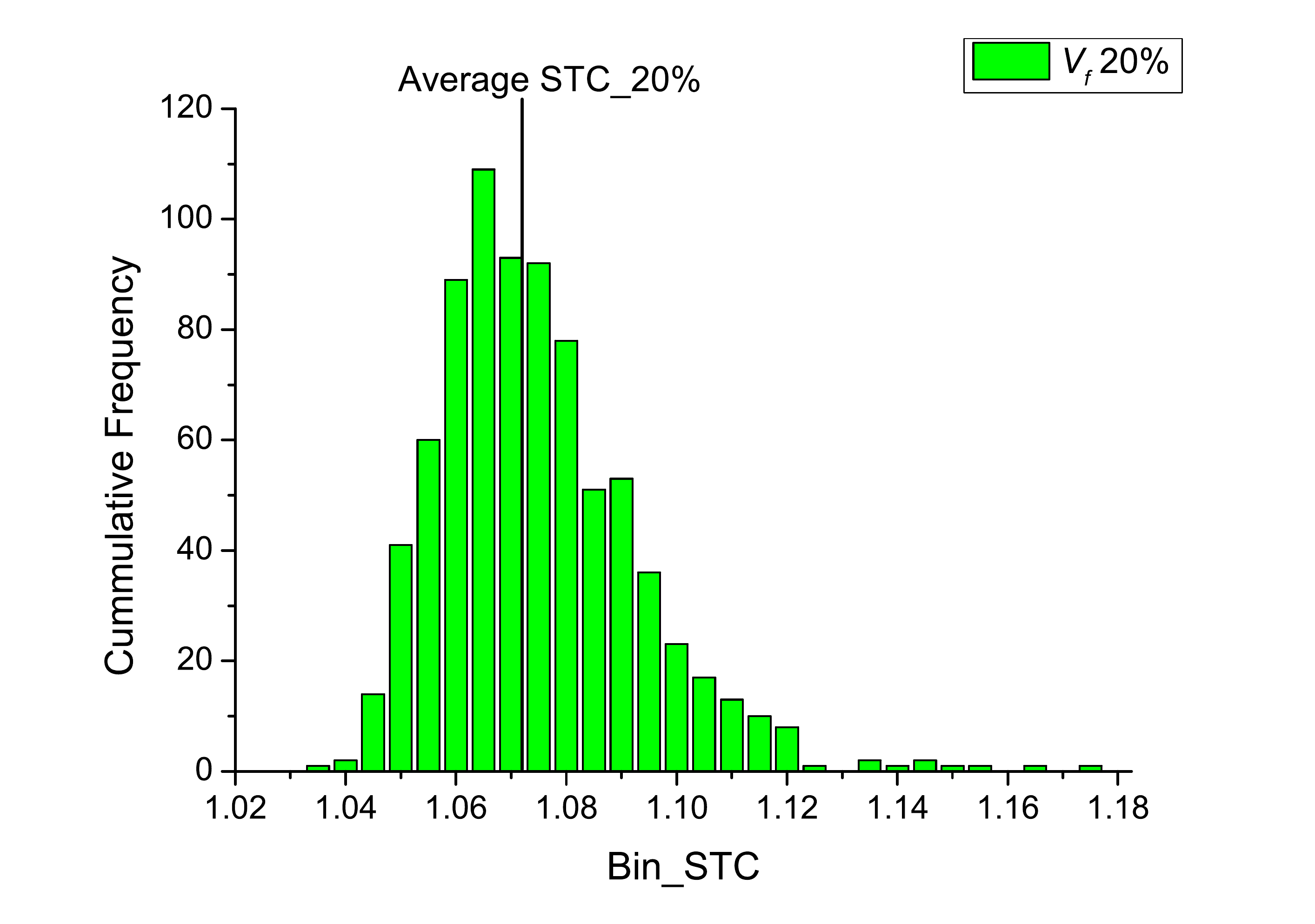}
	}
	~
	\subfigure[]{\includegraphics[width=2in]{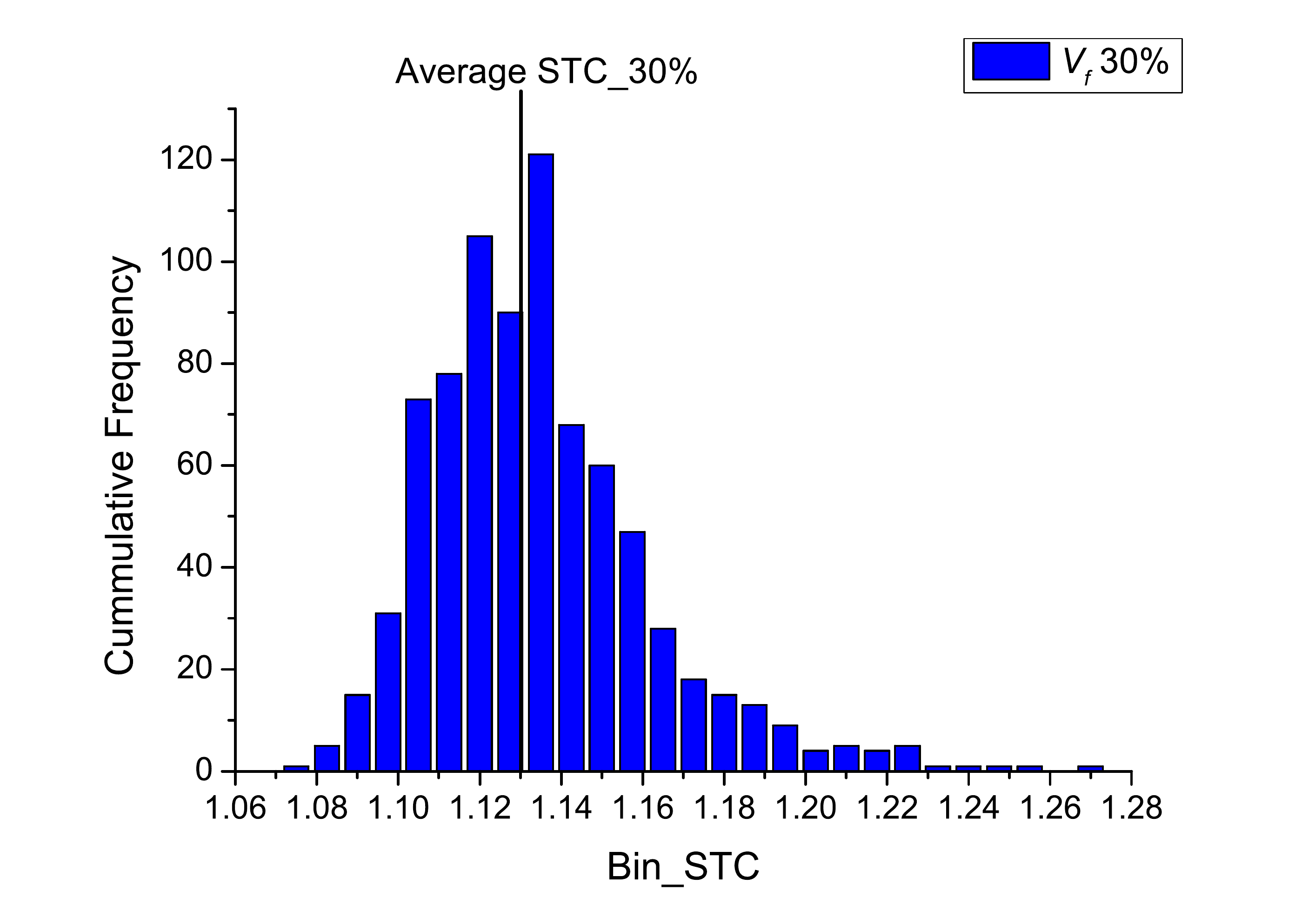}
	}
	~
	\subfigure[]{\includegraphics[width=2in]{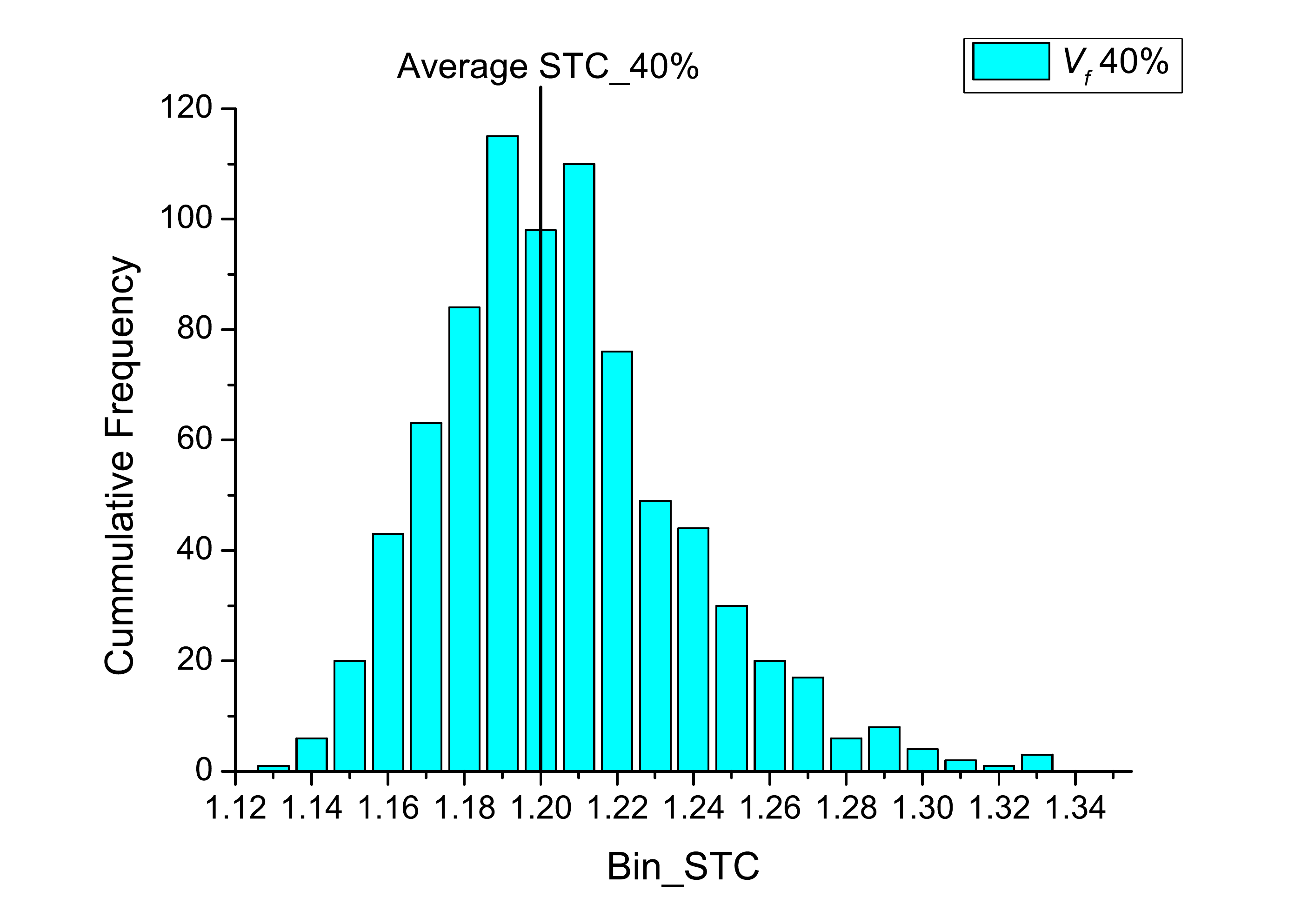}
	}
	~
	\subfigure[]{\includegraphics[width=2in]{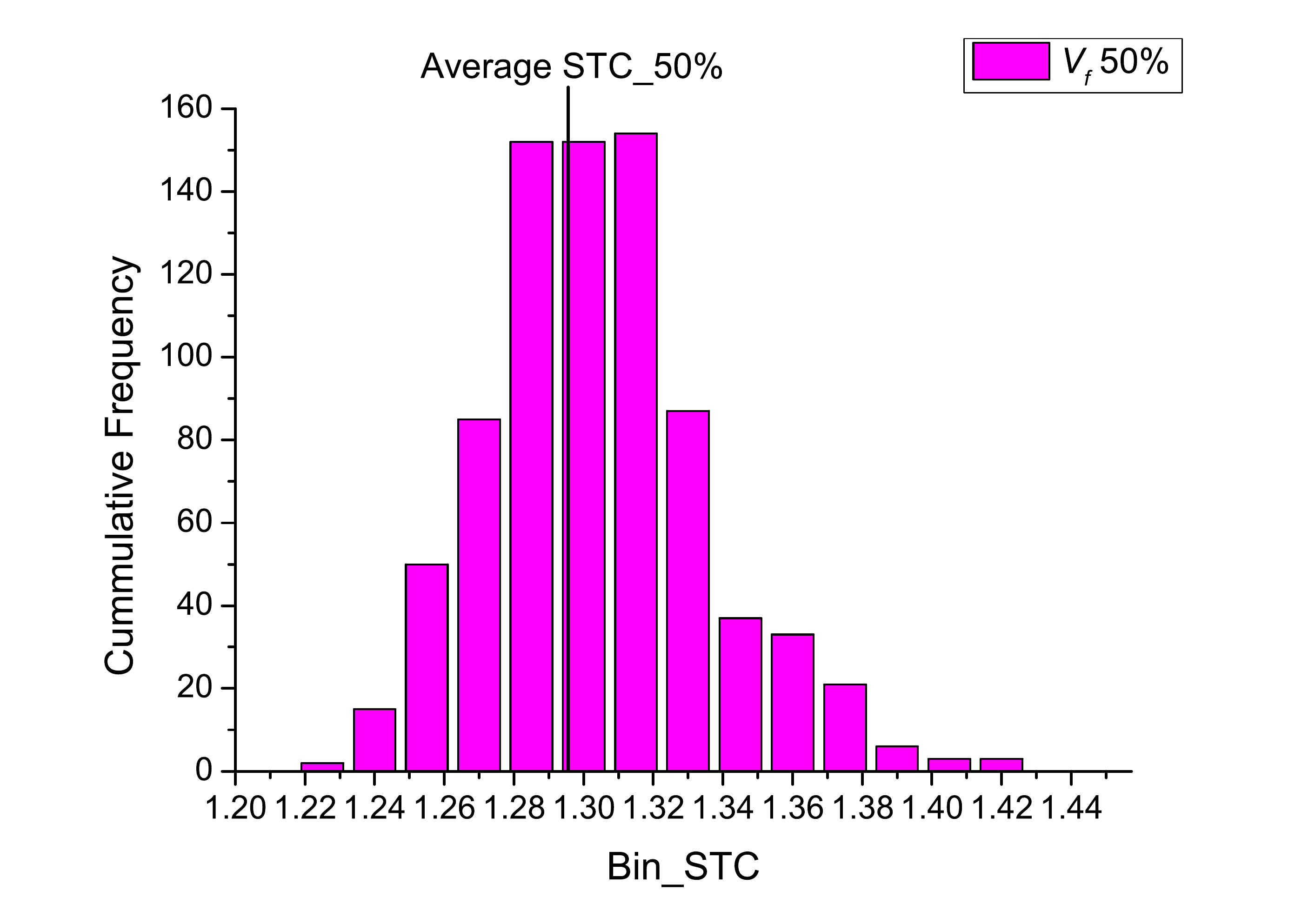}
	}
	~
	\subfigure[]{\includegraphics[width=2in]{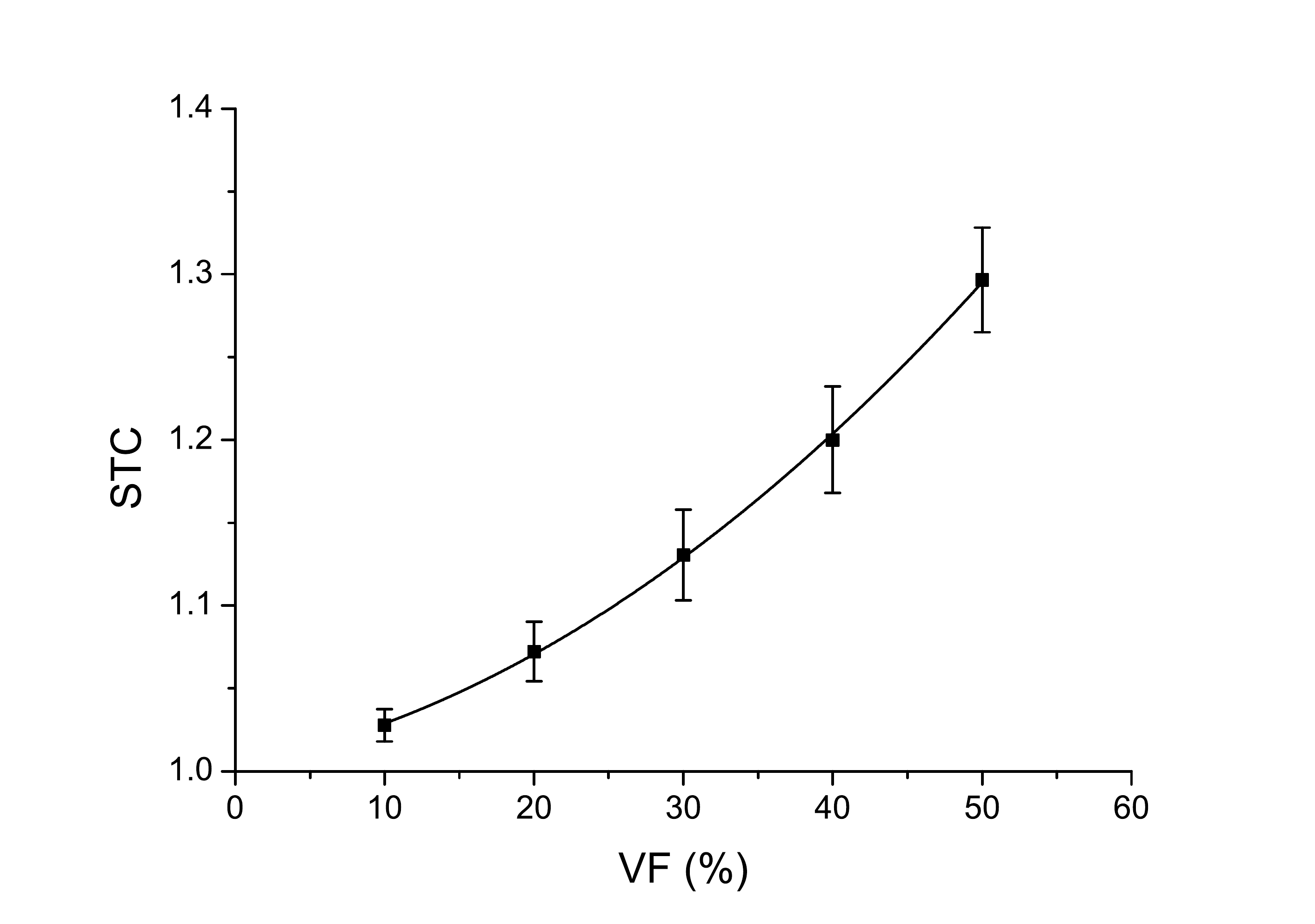}
	}
	\caption{Histogram plot of STC distribution for Case 2. (a) $V_f$ $10\%$, (b)  $V_f$ $20\%$, (c)  $V_f$ $30\%$, (d)  $V_f$ $40\%$, (e)  $V_f$ $50\%$, (f) Variation of average STC along with the standard deviation with respect to volume fraction.}
	\label{fig:fig11}
\end{figure*}

As mentioned earlier the volume fraction, in this case, is changed by changing the size of the RVE keeping the number of fibers and radius of the fibers constant. First, the case of semi-random distribution of fibers is studied for the effect of fiber distribution and fiber volume fraction on STC, and then the special case of periodic RVE is analysed. Up to $50\%$ volume fraction is considered in the semi-random case as the standard deviation of the fiber distribution for $60\%$ and $70\%$ volume fraction is very small, as shown in Table~\ref{tab:tab1}. As stated earlier 200 RVE’s for each volume fraction is generated in which the fibers in the unit cells except central broken fiber are randomly displaced from its mean position but within the unit cells. For the analysis the STC of the nearest fibers (fibers at location 1 in Fig.~\ref{fig:fig8}) are clubbed together since the distance of these fibers at their mean position from the broken fiber is the same. Thus, for each volume fraction, the number of data points of STC is 800. The standard deviation of these 800 STC values is calculated and plotted against volume fraction as shown in Fig.~\ref{fig:fig6a} It is observed that with the volume fraction the standard deviation of the STC distribution increases from $10\%$ to $40\%$ and there is a slight decrease in its value for $50\%$ volume fraction. Fig.~\ref{fig:fig11} shows the histogram plot of STC distribution.

\begin{figure}
	\centering
	\includegraphics[width=3.3in]{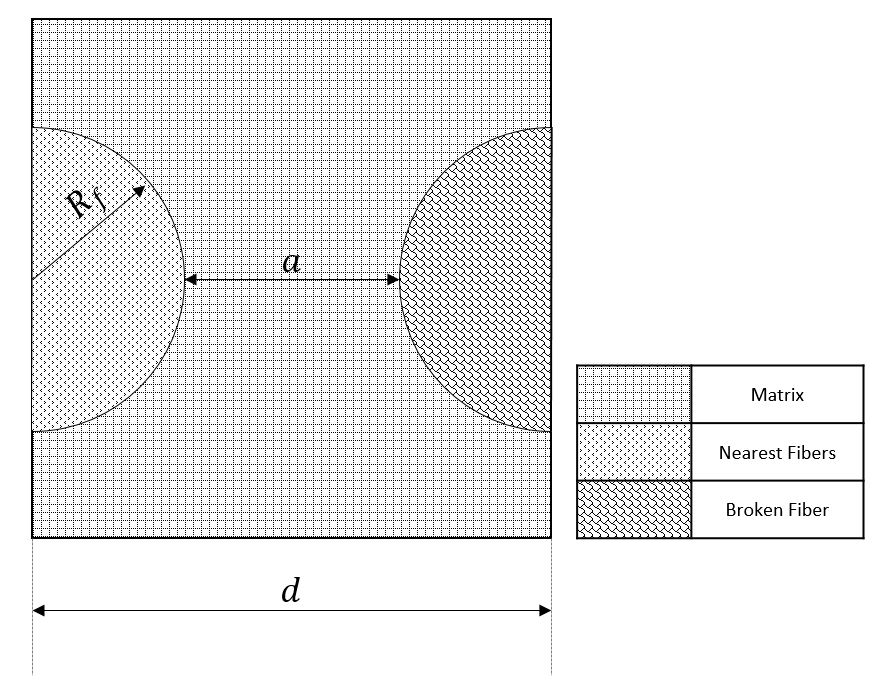}
	\caption{The interaction between two fibers through the matrix.}
	\label{fig:fig12}
\end{figure}

As there are a total of 25 number of fibers, there will fibers in all unit cells (locations 1 to 5 in Fig.~\ref{fig:fig8}) in each RVE. The distance between the surfaces of the broken fiber and the nearest fibers (denoted by ‘a’ in Fig.~\ref{fig:fig12}) dictates the trend observed. Considering the fibers at their mean position, the mean value of ‘a’ is changed by changing the value of ‘d’ ~$(d=\frac{D}{5})$ the size of the unit cell. Further, the value of 'a' is changed by displacing the fibers from their mean position. The value of ‘a’ signifies the amount of matrix present in between the fibers. For a low volume fraction of $10 \%$, even the minimum value of ‘a’ after displacing the fibers at location 1 is so large, implying a large volume of the matrix is present in between the fibers. A large amount of stress transferred from the broken fiber is depleted in the shearing of the matrix and only a small amount of stress is eventually transferred to the nearest fibers. Hence the value of STC is low. The variation in the value of STC in different RVE’s of the same volume fraction is due to different values of ‘a’ of the nearest fibers. Different values of ‘a’ imply that the stress transferred to the nearest fibers is different. It is similar to the shielding effect. The value of ‘a’ for other fibers also contributes to the variation of STC but the nearest fibers value dominates. For a $10 \%$ fiber volume fraction, as the mean value of ‘a’ is very large the changes in its value due to displacement of the fibers from their mean position is negligible and hence the variation in the values of STC is less resulting in the low standard deviation of STC distribution.

As the fiber volume fraction is increasing with the decrease in the value of ‘d’, the mean value of ‘a’ is decreasing implying more stress is transferred to the nearest fibers and less stress is depleted in the shearing of the matrix. Thus, the value of STC is increasing. The change in the mean value of ‘a’ results in the variation of STC values for different RVE’s of the same volume fraction. As the volume fraction is increasing, stress transferred is shared by the nearest fibers instead of shearing the matrix. The standard deviation of STC distribution is thus increasing with the volume fraction. However, after a volume fraction of $40\%$ i.e. for volume fraction $50\%$, there is a small decrease in the value of standard deviation. The reason for this can be seen from Table~\ref{tab:tab1} from which it is clear that the value in the standard deviation of fiber distribution is low, which implies that the RVE is tending towards the periodicity. There is not much variation in the mean value of ‘a’ and hence the variation in STC value is less resulting in a decrease in the value of the standard deviation of STC distribution.

The average value of the STC of these 800 data points of each volume fraction is plotted against the corresponding fiber volume fraction as shown in Fig.~\ref{fig:fig9b}.  From the above discussion, it is seen that for low volume fraction the standard deviation of STC distribution is low which implies that the minimum and the maximum value of STC are close. With the volume fraction as the standard deviation is increasing the min and max values are moving far from each other. Both the min and max values show an increasing trend with volume fraction as the value of ‘a’ is decreasing and hence the average STC value is increasing with the volume fraction. The periodic case is taken as the special case of the semi-random case in which the fibers are not displaced from their mean position. In this case, the value of ‘a’ is changed only because of the variation in the size of the RVE. The values of STC for the periodic case are found to coincide with the average value of STC in the semi-random case as shown in Fig.~\ref{fig:fig9b}.

The average ineffective length decreases with the increases in the volume fraction similar to Case 1. The plot of the average ineffective length for different volume fractions is shown in Fig.~\ref{fig:fig10b}. The standard deviation of the values of the ineffective length plotted concerning the fiber volume fraction is shown in Fig.~\ref{fig:fig6b}. From the figure, it is observed that as the volume fraction increases the standard deviation of the distribution of the ineffective length increases up to $40\%$ starting from zero at fiber volume fraction $10\%$, then it suddenly decreases again to zero for fiber volume fraction 50 and $60\%$. From the scenario of one and more than one intact fibers, it is clear that smaller ineffective length is observed when the intact fibers are very near to the broken fiber and also when there are more number of fibers surrounding the broken fiber. However, as in this case the number of fibers surrounding the broken fiber are the same in all RVE’s of all volume fraction i.e. 24. All the 24 locations mentioned in Fig.~\ref{fig:fig8}. are occupied by these 24 intact fibers. Now the only parameter which is responsible for the observed variation in the ineffective length is the distance between the broken fibers and the intact fibers (the value of ‘a’ shown in Fig.~\ref{fig:fig12}).

The mean value of ‘a’ for the fiber volume fraction of $10\%$ is very large. Even the minimum value due to the repositioning of the fibers is very large. This means that there is a large volume of the matrix surrounding the fiber, implying large shearing of the matrix. Thus the observed value of the average ineffective length is large among all the volume fractions. As mentioned earlier that the standard deviation of the distribution of the ineffective length is zero for the fiber volume fraction of $10\%$ which implies that all the RVE’s results in the same value of ineffective length. The mean value of ‘a’ being very large, the changes in its value especially for $1^{st}$ and $2^{nd}$ nearest fibers will affect the ineffective length negligibly. The ineffective length obtained from the algorithm is the rounded value as stated earlier, and as the variation in ineffective length observed due to changes in the value of ‘a’ is negligible, a constant value of ineffective length is observed in all RVE’s resulting in zero value of standard deviation. As the volume fraction is increasing with the decrease in the size of the unit cell (value of ‘d’ in Fig.~\ref{fig:fig12}), the mean value of ‘a’ is decreasing. As the mean distance between the fibers is decreasing the matrix present between the fibers is decreasing, implying less shearing of the matrix. The intact fibers restrict the displacement of the broken fiber, helping it building the stress. The average value of ineffective length thus decreases with volume fraction.  As the fibers are repositioning from their mean positions in different RVE’s the mean value of ‘a’ for $1^{st}$ and $2^{nd}$ nearest fibers is changing, (the value of ‘a’ is changing for all the fibers but the $3^{rd}$ $4^{th}$ and $5^{th}$ nearest fibers contribute very less in the rebuilding of the stress in the broken fiber) resulting in the variation of the values of the ineffective length implying higher values of the standard deviation. The highest value of standard deviation is observed for volume fraction $40\%$.

With further increase in the volume fraction, the average value of ineffective length decreases due to less shearing of the matrix. However, the value of the standard deviation of ineffective length decreases to zero for the fiber volume fraction of $50\%$ and $60\%$. This is because the value of ‘d’ the size of the unit cell becomes small with the increase in the volume fraction. The standard deviation of the distribution of fibers in the unit cell is very small, as shown in Table~\ref{tab:tab1} implying negligible variation in the mean value of ‘a’ and hence less variation in the values of the ineffective length. The RVE’s are becoming more periodic and hence resulting in constant values of ineffective length making the standard deviation to zero. The plot for the ineffective lengths for the volume fractions having all the fibers at their mean positions (square arrangement) is also shown in Fig.~\ref{fig:fig10b} It is observed that like STC the values of the average ineffective length matches with the values from complete periodic arrangements of the fibers.

\subsection{Analysis of Case 3}
\begin{figure*} 
		\centering
	\subfigure[]{\includegraphics[width=2in]{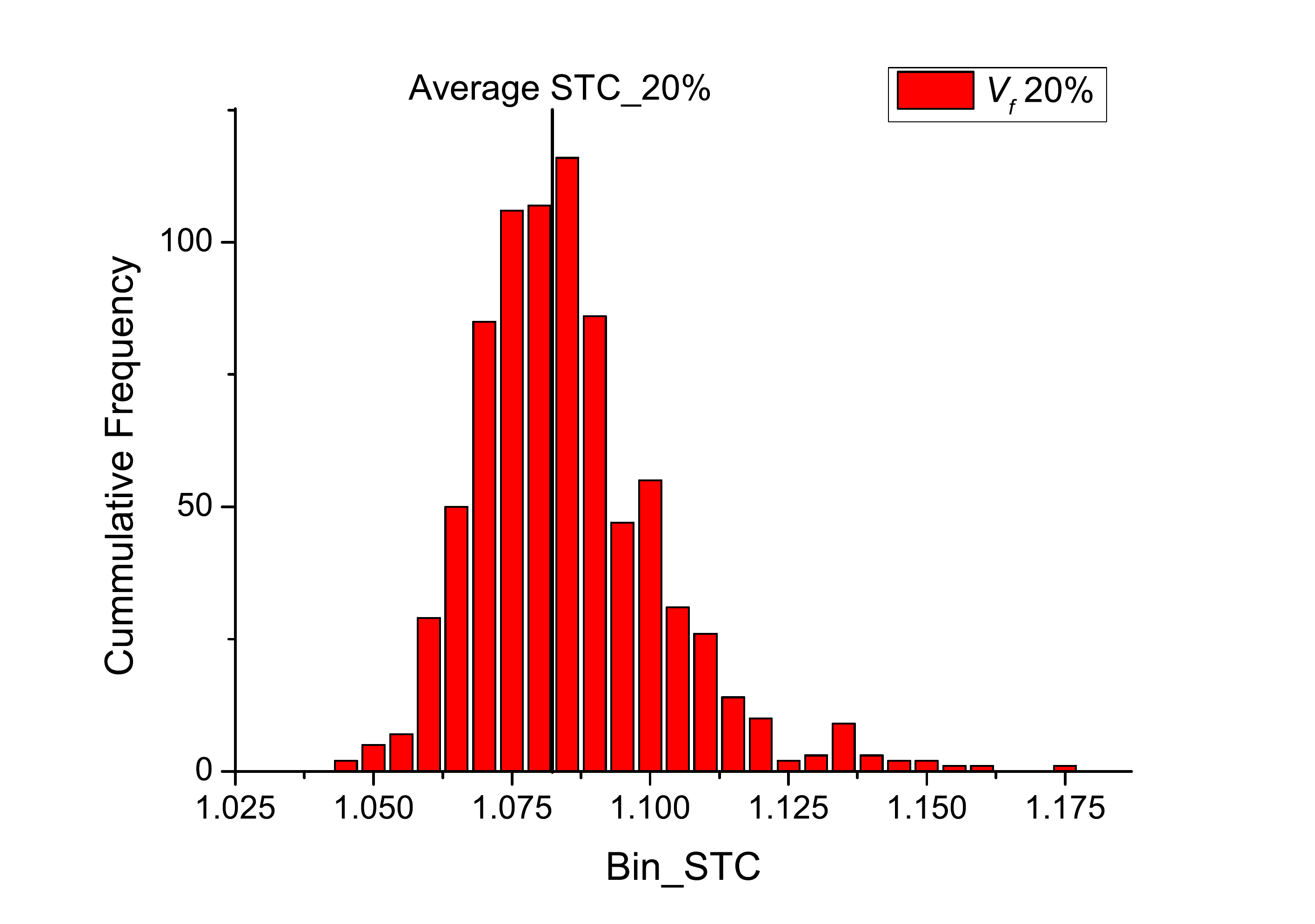}
	}
	~
	\subfigure[]{\includegraphics[width=2in]{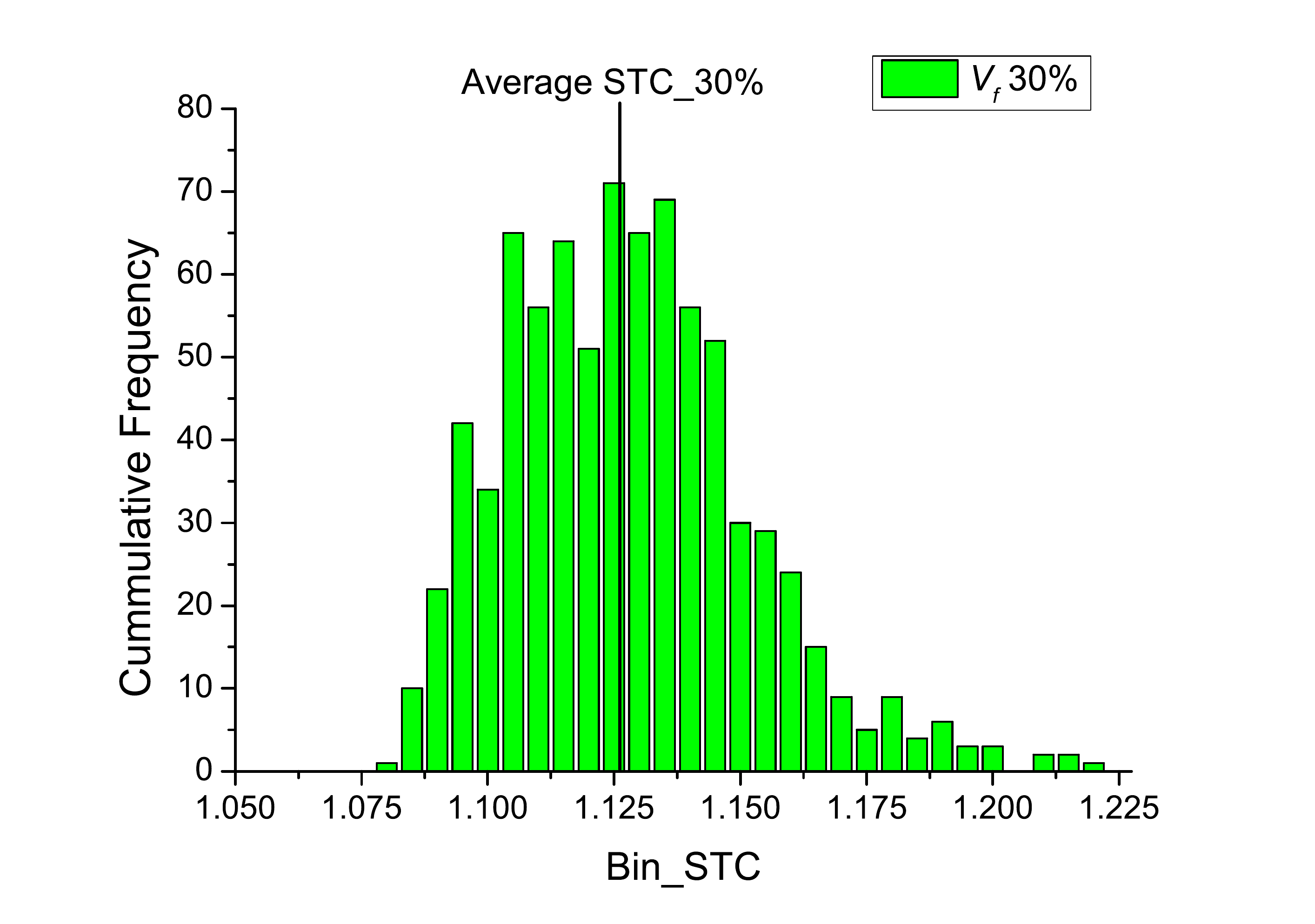}
	}
	~
	\subfigure[]{\includegraphics[width=2in]{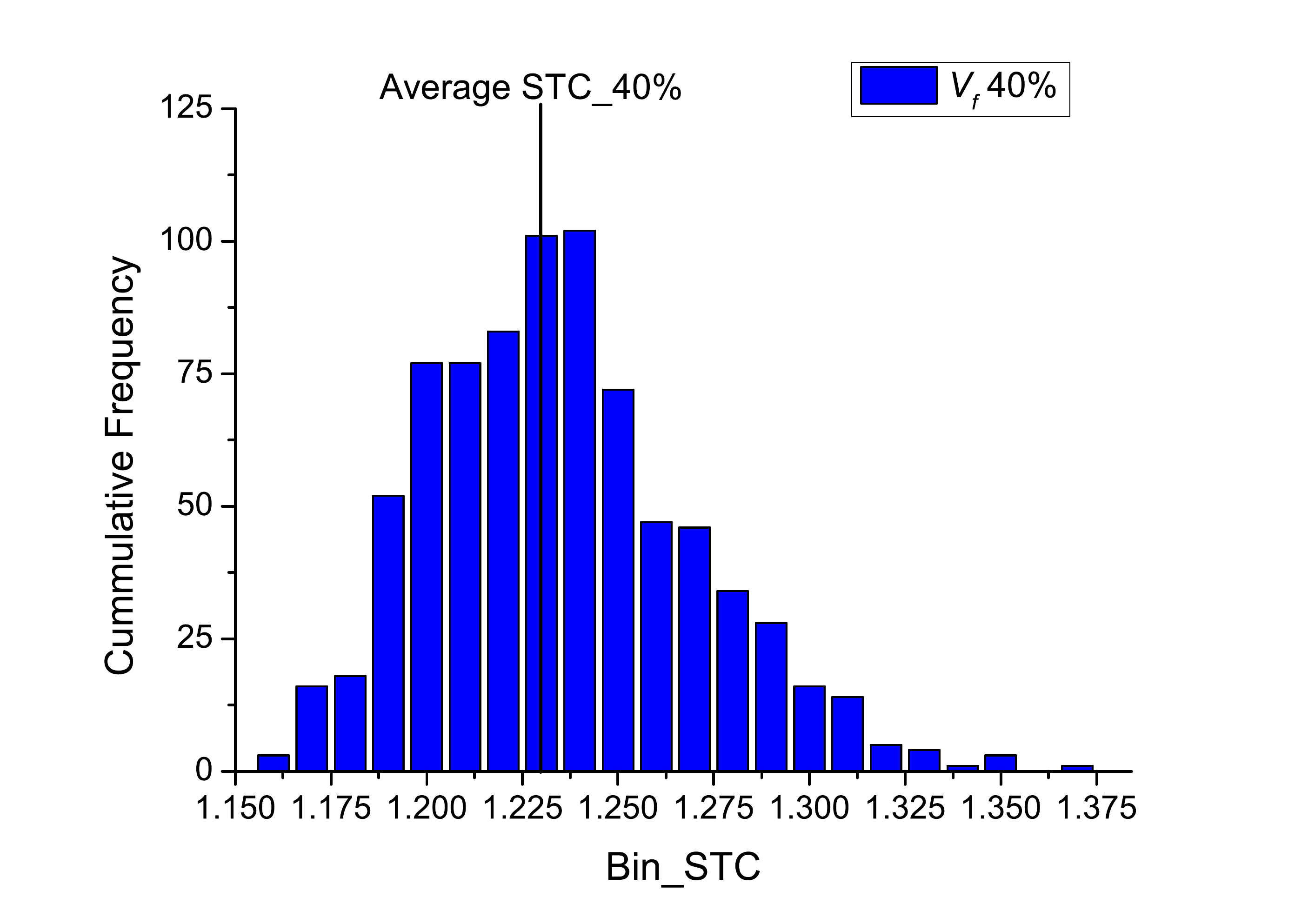}
	}
	~
	\subfigure[]{\includegraphics[width=2in]{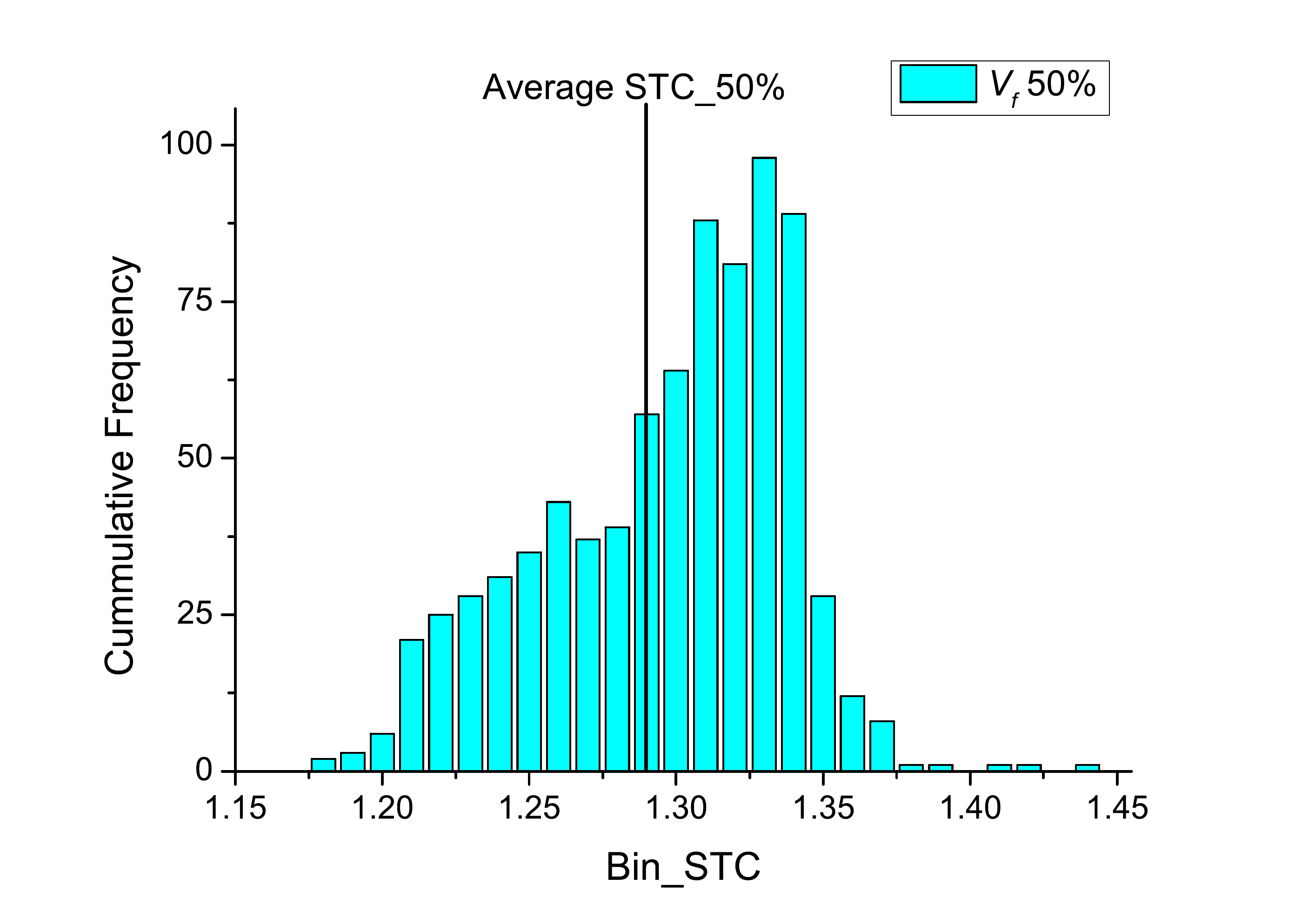}
	}
	~
	\subfigure[]{\includegraphics[width=2in]{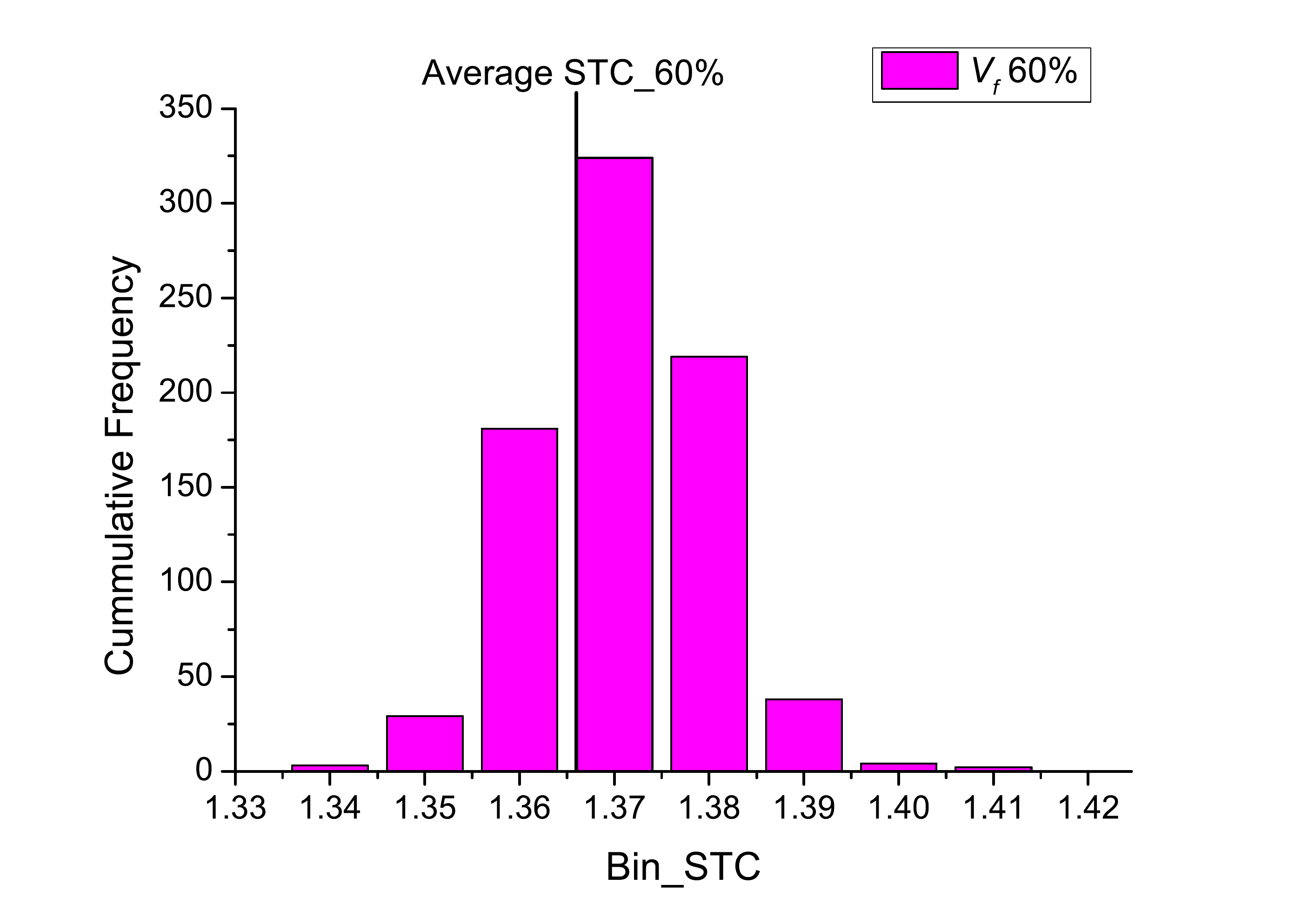}
	}
	~
	\subfigure[]{\includegraphics[width=2in]{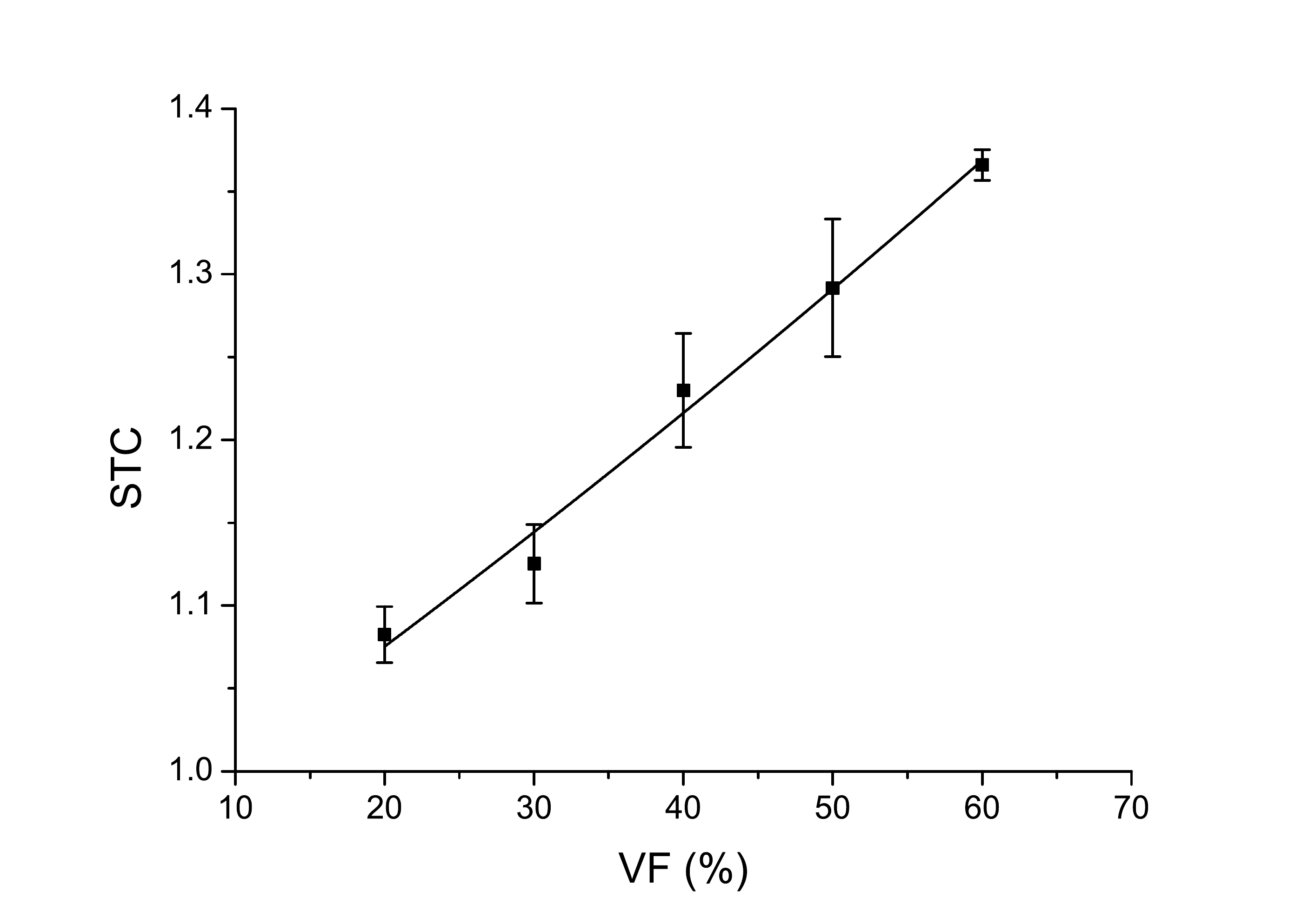}
	}
	\caption{Histogram plot of STC distribution for Case 3. (a) $V_f$ $10\%$, (b)  $V_f$ $20\%$, (c)  $V_f$ $30\%$, (d)  $V_f$ $40\%$, (e)  $V_f$ $50\%$, (f) Variation of average STC along with the standard deviation with respect to volume fraction.}
	\label{fig:fig13}
\end{figure*}
In this case, the volume fraction is changed by changing the radius of the fibers keeping the number of fibers and size of the RVE constant. For the semi-random case, $20\%- 60\%$ of volume fraction is considered as the standard deviation of the fiber distribution for $70\%$ volume fraction is very small, as shown in Table~\ref{tab:tab1}. In this case, also 800 data points of STC are generated for each volume fraction from 200 RVE’s and clubbing the nearest fiber STC data together. The standard deviation of these 800 STC values is calculated and plotted against volume fraction, as shown in Fig.~\ref{fig:fig6a}. A histogram plot of the STC distribution is shown in Fig.~\ref{fig:fig13}. Similar to Case 2 it is observed that with the volume fraction the standard deviation of the STC distribution increases from $20\%$ to $50\%$ and there is a decrease in its value for $60\%$ volume fraction. Fig.~\ref{fig:fig9c} shows the variation of average, maximum and minimum values of STC with the increase in fiber volume fraction.

The reason for this trend remains the same as that for Case 2, the means of changing the  value of ‘a’ is different. It is varied by changing the fiber radius ‘$R_f$’ while ‘d’ in this case remains constant. Also, it is observed that in this case also the values of STC for periodic RVE approximately match with the average STC values for the corresponding fiber volume fraction, as shown in Fig.~\ref{fig:fig9c}. It is to be noted that the maximum value of STC for a $60\%$ volume fraction is less than the $50\%$ volume fraction. The reason for it is that the standard deviation of the fiber distribution in the $60\%$ volume fraction is very less than $50\%$. The drop in the value of the standard deviation of fiber distribution from $50\%$ to $60\%$ volume fraction is almost $91\%$. Due to which all the $1^{st}$ nearest fibers, as shown in Fig.~\ref{fig:fig8} are approximately at their mean positions and hence equal stress is shared by all the four fibers resulting in a low value of the STC.

\begin{figure}
	\centering
	\includegraphics[width=3.3in]{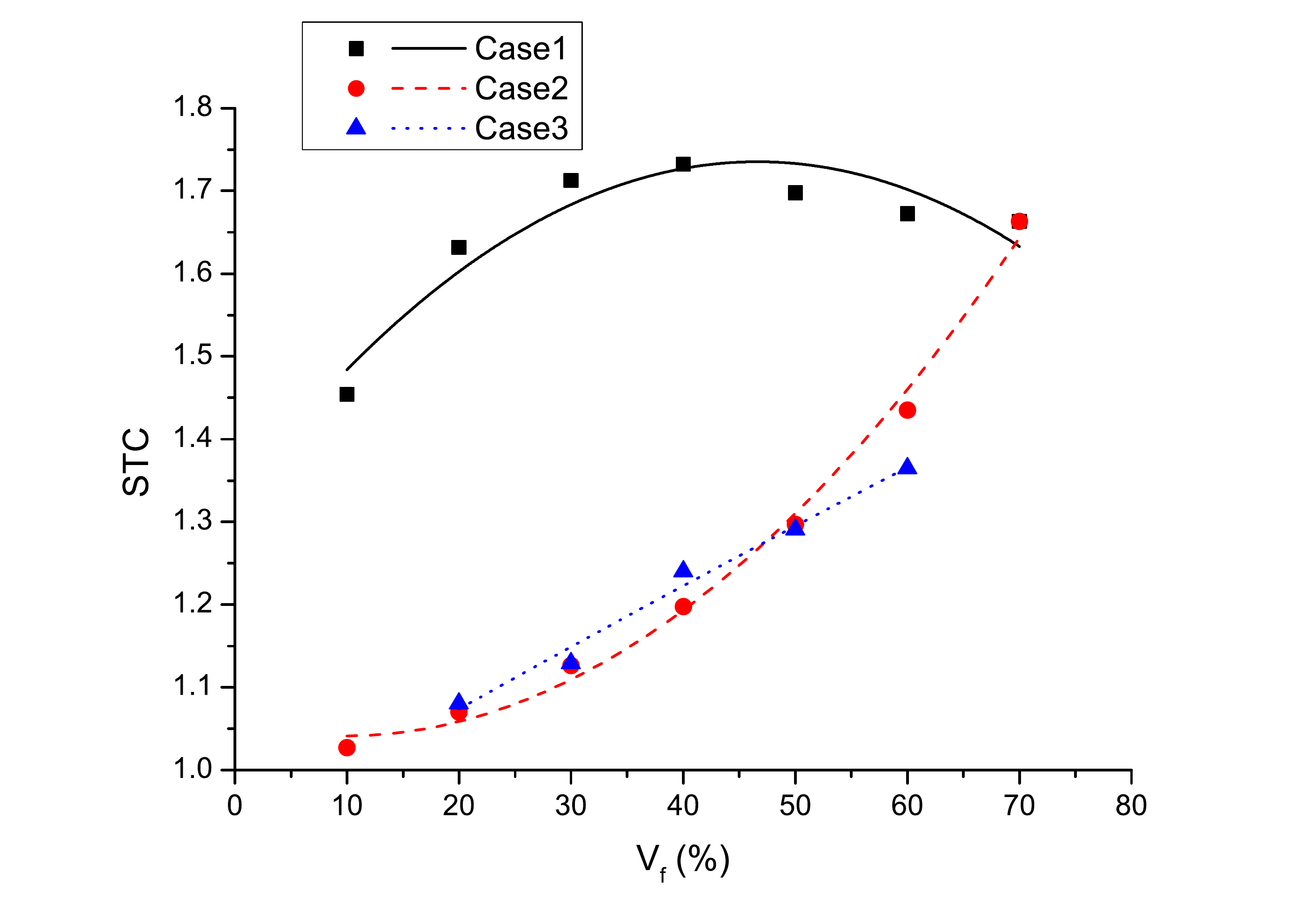}
	\caption{Comparison of the values of the average STC in all cases 1, 2, and 3.}
	\label{fig:fig14}
\end{figure}
The variation of the standard deviation of the values of ineffective length is shown in Fig.~\ref{fig:fig6b}. Interestingly the result of average values of ineffective length and also the ineffective length of RVE’s of all-fiber volume fraction having fibers at their mean position (periodic arrangement) shows a reverse trend compared with Case 1 and Case 2. The value of ineffective length is observed to increase with the volume fraction, as shown in Fig.~\ref{fig:fig10c}. The fiber radius is the reason for the observed trend. For low volume fraction, the radius of the fiber is too small. Less amount of shearing of the matrix will suffice the rebuilding of the strength of the broken fiber. The mean value of ‘a’ being large, the effect of fiber distribution on the ineffective length is not observed and hence the value of the standard deviation of the ineffective length is zero for fiber volume fraction $10\%-30\%$. This value rises in $40\%$ and then again approaches zero at $50\%$ and becomes zero at $60\%$ suggesting the RVE’s are tending towards periodic arrangement since the values of the standard deviation of fiber distribution are very low. The average value of ineffective lengths, when compared with the values for periodic arrangement, shows closeness in them. From the above observations, it can be said that as the ratio of $\frac{l}{R_f}$ decreases the ineffective length increases\footnote{l being constant equal to $100~\mu m$.}. The small the radius of the fiber quicker will be the gain in the strength.
Comparing the average STC values in all three cases it is found that the STC values in Case 1 are much more than Case 2 and 3. Case 2 and Case 3 has almost similar values up to $50\%$, as shown in Fig.~\ref{fig:fig14}. The average ineffective length is observed to be maximum in Case 2.

\section{Conclusion}\label{sec:conclusion}
A thorough analysis has been done to study the effect of volume fraction and fiber distribution on the STC and the ineffective length. To the best of author’s knowledge, this is a first attempt comparing the effects of volume fraction on STC and ineffective length in three different cases of fiber volume fraction variation has been performed. The three cases considered were Case 1: change in the number of fibers, Case 2: considering change in the dimension of the RVE, and Case 3: change in the fiber radius. The results obtained and discussed in all three cases suggest that the trends of the distribution of STC and ineffective length for 200 RVE’s for each fiber volume fraction, can be generalized for the same loading conditions with different numerical values.

From Case 1 and semi-random fiber distribution of Case 2 and Case 3, it is observed that even though the fiber arrangement follows the normal distribution, the distribution of STC does not follow any standard distribution. A similar kind of observation was observed by~\cite{Fukuda1985}. From Case 1, it is observed that the standard deviation of STC decreases with the increase in the fiber volume fraction, implying that the RVE tends towards periodicity with the volume fraction. As stated earlier, that the fibers should be ideally placed in the periodic arrangement, from the standard deviation this can be confirmed. In high fiber volume fraction, the deviation of the STC from the average is very minimal and hence the probability of the fibers failing due to an increase in the stress is small. Whereas, in the low fiber volume fraction, even though the average STC is less, but it is to be noted that the maximum value of the STC among all the volume fractions is observed in this volume fraction, indicating a high risk of fiber failure. There is a large amount of uncertainty in the low volume fraction composites. With the increase in the fiber volume fraction this uncertainty decreases. The average ineffective length is found to be minimum in the highest fiber volume fraction. This indicates that the shielding effect plays a very vital role in regaining the strength of the broken fiber.

From Case 2 and Case 3 it is observed that the semi-random arrangement of fibers does not provide any significant benefit in micromechanical analysis over the periodic arrangement as the average values of the STC’s and the average values of the ineffective length in both the arrangements are almost the same. In Case 2 and Case 3 as all the 25 unit cells are always occupied by the fibers, the shielding effect due to the intact fibers (assuming all the fibers at their mean position) is the same. The variation in the values of the STC and ineffective length in the semi-random case is due to the change in the average distance between the broken fiber and the nearest fibers. The effect of the change in the average distance from the broken fiber and second, third fourth, and fifth nearest fibers decreases respectively. From Case 3 it is observed that the ineffective length of the fiber depends on the cross-sectional dimension of the fiber. The smaller the radius of the fiber smaller is the ineffective length. 
Considering the perfect fiber-matrix bond and the same fiber dimensions, for random fiber distribution (Case 1) higher fiber volume fraction FRC is advisable.  Case 2 suggests that the fiber volume fraction $40\%$ to $50\%$ is the optimum volume fraction considering the trends of the average STC and average ineffective length. While Case 3 suggests that low fiber volume fraction is the suggested, since in low volume fraction the average STC and average ineffective length is the minimum. Comparing all three cases, the average STC is found to be high in the random arrangement. Hence, while manufacturing the FRC’s care should be taken to minimize the error to maintain the periodic arrangement of the fibers.

\bibliographystyle{unsrt}  
\bibliography{Literature}  


\end{document}